\documentclass[a4paper,11pt]{article}

\usepackage[english]{babel}
\usepackage{amssymb}
\usepackage{amsmath}
\usepackage{amsthm}
\usepackage{psfrag}
\usepackage[T1]{fontenc}
\usepackage{ae,aecompl}
\usepackage[colorlinks]{hyperref}
\usepackage{subfigure}
\usepackage{appendix}
\usepackage{natbib}
\hypersetup{colorlinks,breaklinks=true,citecolor=blue,linkcolor=blue}
\usepackage{graphicx}
\usepackage{url}

\providecommand\bnabla{\boldsymbol{\nabla}}
\providecommand\bcdot{\boldsymbol{\cdot}}

\newcommand\Rey{\mbox{\textit{Re}}}  
\newcommand\Pran{\mbox{\textit{Pr}}} 

\newcommand\eg{e.g.\ }
\newcommand\ie{i.e.\ }

\newcommand{\pleft}{\left(}
\newcommand{\pright}{\right)}
\newcommand{\vect}[1]{\boldsymbol{#1}}

\newcommand{\vel}{\boldsymbol{u}}

\newcommand{\vor}{\boldsymbol{\omega}}
\newcommand{\pdt}[1]{\frac{\partial #1}{\partial t}}
\newcommand{\Ek}{\mbox{\textit{Ek}}}
\newcommand{\Ra}{\mbox{\textit{Ra}}}
\newcommand{\tRa}{\widetilde{\textit{Ra}}}

\newcommand{\Ro}{\mbox{\textit{Ro}}}
\newcommand{\Nu}{\mbox{\textit{Nu}}}

\newcommand{\aspect}{\lambda}
\newcommand{\RNu}{\textit{R}_\textit{Nu}}

\title{Large-scale vortices in rapidly rotating Rayleigh-B\'enard convection}

\author{C\'eline Guervilly, David W. Hughes \& Chris A. Jones \vspace{0.2cm}
 \\ {\small Department of Applied Mathematics,}
 \\ {\small University of Leeds, Leeds LS2 9JT, UK}}
\begin{document}

\maketitle

\begin{abstract}
Using numerical simulations of rapidly rotating Boussinesq convection in a Cartesian box, 
we study the formation of long-lived, large-scale, depth-invariant coherent structures. 
These structures, which consist of concentrated cyclones, grow to the horizontal scale of the box, 
with velocities significantly larger than the convective motions. We vary the rotation rate, the thermal 
driving and the aspect ratio in order to determine the domain of existence of these large-scale vortices (LSV). 
We find that two conditions are required for their formation. First, the Rayleigh number, a measure of the thermal driving, 
must be several times its value at the linear onset of convection; this corresponds to Reynolds numbers, 
based on the convective velocity and the box depth, $\gtrsim 100$. Second, the rotational constraint on the 
convective structures must be strong. This requires that the local Rossby number, based on the convective velocity 
and the horizontal convective scale, $\lesssim 0.15$. Simulations in which certain wavenumbers are artificially 
suppressed in spectral space suggest that the LSV are produced by the interactions of small-scale, depth-dependent 
convective motions. The presence of LSV significantly reduces the efficiency of the convective heat transport.
\end{abstract}

\section{Introduction}
The presence of large-scale coherent structures in turbulent flows attracts much interest, 
particularly because of their relevance in geophysics and astrophysics; understanding their formation is 
thus an important problem in fluid dynamics. In two-dimensional (2D) turbulence, in which vertical motions are 
assumed to be suppressed owing to strong stratification, fast rotation, or small vertical to horizontal scale ratio, 
the kinetic energy and the enstrophy (mean-square vorticity) are conserved quantities in the inviscid limit. 
This implies a downscale enstrophy cascade and an upscale energy cascade \citep{Kra67}, which can lead to the 
formation of coherent vortices \citep[e.g.][]{mcwilliams1984}. By contrast, in three-dimensional (3D) isotropic 
turbulence, enstrophy is not a conserved quantity, and the energy cascade is expected to be downscale. Nevertheless, 
the constraint imposed by rapid rotation might lead to 2D-like dynamics in a 3D flow on timescales longer than 
the rotation period. Notable examples of the formation of coherent vortices in a 3D system subject to rotation have 
been observed in experiments on grid-generated turbulence \citep[e.g.][]{Hopfinger1982, Staplehurst2008}. Recently, 
the presence of large-scale vortices (LSV) has been reported in numerical simulations of rotating convection in Cartesian 
geometry \citep{Chan07}, where the fluid is heated from below and confined between two horizontal planes. 
In this system, with buoyant vertical motions, the flow is necessarily $z$-dependent. The formation of LSV in 
convective layers still remains to be fully understood. 

Using numerical models of rotating compressible thermal convection in a local $f$-plane model, \citet{Chan07} and 
\citet{Chan13} report the emergence of long-lived, large-scale (\ie domain size) vortices for sufficiently large 
rotation rates. For moderate Rossby number ($\Ro$, the ratio of the rotation period to the typical convective 
turnover time), of the order of $0.1$, these LSV are cyclonic and associated with regions of lower temperature 
relative to their surroundings. Note that a vortex is defined as cyclonic (anticyclonic) when its vorticity 
in the rotating reference frame has the same (opposite) sign as the externally applied rotation. At lower Rossby 
numbers, \citeauthor{Chan07} and \citeauthor{Chan13} observe a large-scale warmer anticyclone accompanied by a 
smaller and weaker cyclone. Using a similar numerical set-up, \citet{Kap11} find that the LSV are excited provided 
that the Reynolds number ($\Rey$, the ratio of the viscous diffusion time to the convective turnover time) is 
sufficiently large. The vortices span the entire vertical extent of the box and are roughly aligned with the 
rotation axis. \citet{Man11} attribute the formation of these structures to a mean-field hydrodynamical 
instability that requires a sufficient scale separation between the convective eddies and the smallest 
horizontal wavenumber permitted in the computational domain. They find that increasing the box size leads to an 
increase of the horizontal extent of the structures, so that the LSV always fill roughly half of the horizontal domain.

Large-scale structures have also been described in the work of \citet{Jul12}, who employ a set of reduced equations 
in a local Cartesian box describing Boussinesq convection in the limit of small Rossby number. In their model, 
the flow is locally in geostrophic balance at leading order $1/\Ro$, but thermally driven vertical flows exist at 
sufficiently small horizontal scales. When the thermal forcing is sufficiently large, \citeauthor{Jul12} observe the 
formation of a depth-invariant box-size flow, which becomes organised into a cyclone and anticyclone of similar strength. 
Using the same numerical model, \citet{Rubio13} show that the generation of these depth-invariant LSV involves the 
interactions of small-scale, depth-dependent convective eddies, which are made more coherent by the action of 
depth-invariant vortices. Interestingly, \citet{Jul12} find that the presence of LSV tends to increase the efficiency 
of the heat transfer through the system.

Fully 3D Boussinesq convection in the presence of rotation has been extensively studied, particularly in 
cylindrical and spherical geometries with applications to the global dynamics of planetary 
interiors \citep[e.g.][]{Busse1994,Chr02}. In spherical geometry, the curved boundaries have an important effect 
on large-scale structures; in flows with large Reynolds numbers and low Rossby numbers they are, notably, responsible 
for the formation of zonal flows of amplitude large compared with the typical convective velocity \citep[e.g.][]{Hei05}. 
In simulations of rotating convection in spherical geometry, the formation of vortices at scales larger than the typical 
convective size has not been observed. \citet{Jul12} conjecture that, in a Cartesian 
domain, the size of the LSV is limited only by the domain size, so that if the upscale energy transfer were allowed to 
continue, the LSV would eventually feel the latitudinal variation of the Coriolis parameter. In this case, it is 
argued that the large-scale dynamics would become organised into zonal flows. 
That said, it is worth highlighting the occurrence of planetary polar vortices --- most strikingly 
those of Saturn --- the dynamics of which may be related to the dynamics of LSV in plane layer models.

In numerical modelling, computational resources limit the values of parameters such as the Reynolds and Rossby numbers; 
this is even more pronounced in global spherical models compared with those in Cartesian geometry. Consequently, 
studies that aim to determine transitions between different convection regimes across a wide parameter range preferentially 
employ the rotating Rayleigh-B\'enard (RRB) configuration, in which a Boussinesq fluid contained between 
two horizontal planes rotates uniformly about an axis aligned with the direction of gravity 
\citep[e.g.][]{Julien1996, Vorobieff2002}. To our knowledge, among the previous studies of RRB convection conducted 
in the low Rossby number regime \citep{King2012, SchTil09, Stellmach2004}, the formation of box-size, 
vertically aligned vortices is addressed only in the contemporaneous study of \citet{Favier2014}. There are 
two possible explanations for the absence of LSV in most of the previous studies. One  stems from the choice 
of boundary conditions, especially for the velocity. In the compressible convection models mentioned above, 
and also in the reduced Boussinesq model of \citet{Jul12}, stress-free boundary conditions are employed; 
often though, no-slip boundary conditions are adopted in RRB convection models \citep[e.g.][]{SchTil10, King2012}. 
Another plausible explanation for the lack of LSV stems from the choice of aspect ratio of the computational domain. 
In RRB simulations, the aspect ratio is usually taken equal to unity or smaller, whereas in the compressible 
convection studies, the aspect ratio is usually about four.

Simulations of RRB convection are often carried out in order to assess the efficiency of heat transfer, 
thereby allowing the determination of the transition between rapidly rotating and non-rotating convection. 
As observed by \citet{Jul12}, heat transfer can be affected by the presence of LSV. It is therefore important 
to identify the conditions required for the formation of the LSV in RRB convection, and to assess their impact on heat transfer. 

In this paper, we investigate in detail the emergence of large-scale, depth-invariant vortices
in convective regions via a series of numerical simulations of RRB convection.
Our objectives are threefold: (i) to determine the parameters governing the presence of LSV; 
(ii) to understand the mechanism by which they form; 
(iii) to assess how LSV affect the heat transfer in the system.

The layout of the paper is as follows. The mathematical and numerical formulation of the problem 
is contained in \S\,\ref{sec:MF}. The formation, maintenance and influence of the LSV are described 
in \S\,\ref{sec:LSV}. The spatial structure of the large-scale vortices, which always consist of a concentrated 
cyclone and a more dilute anticyclone, is discussed in \S\,\ref{sec:structure}, the domain of existence in parameter 
space in \S\,\ref{sec:domain}, and the reasons for the cyclonic/anticyclonic asymmetry in \S\,\ref{sec:cyclonic}. 
In \S\,\ref{sec:filter}, we establish how energy is transferred to the large scales. Finally, in \S\,\ref{sec:heat}, 
we discuss how the LSV affect the heat transfer in the system. A concluding discussion is contained in \S\,\ref{sec:discussion}.

\section{Mathematical Formulation}
\label{sec:MF}

We study rotating Boussinesq convection in a three-dimensional Cartesian domain. 
The motions are driven by an initially uniform temperature gradient, imposed by fixing the temperature on 
the top and bottom boundaries. Acceleration due to gravity is constant, $\vect{g} = - g \vect{e}_z$. 
The rotation vector $\Omega \vect{e}_z$ is aligned with the vertical direction. 
The box depth is $d$. The horizontal dimensions of the computational domain are equal 
in the $x$ and $y$ directions, with the ratio of horizontal to vertical dimensions denoted by $\lambda$.
The fluid has kinematic viscosity $\nu$, thermal diffusivity $\kappa$ and 
thermal expansion coefficient $\alpha$, all of which are constant. We solve the momentum and temperature 
equations in dimensionless form, obtained by scaling lengths with $d$, times with $1/(2\Omega)$, 
and temperature with $\Delta T$, the temperature difference across the layer. In standard notation, 
the complete system of dimensionless governing equations can then be written as
\begin{eqnarray}
	&& \pdt{\vel} + \vel \bcdot \bnabla \vel + \vect{e}_z \times \vel =
	- \bnabla p
	+ \frac{\Ra \Ek^2}{\Pran} \theta \vect{e}_z + \Ek \nabla^2 \vel , 
	\label{eq:u}
	\\
	&& \bnabla \bcdot \vel = 0 ,
	\\
	&& \pdt{\theta} + \vel \bcdot  \bnabla \theta - u_z = \frac{\Ek}{\Pran} \nabla^2 \theta ,
	\label{eq:theta}
\end{eqnarray}
where $\vel=(u_x,u_y,u_z)$ is the velocity field, $p$ the pressure and $\theta$ the temperature perturbation 
relative to a linear background profile. The dimensionless parameters are the Rayleigh number,
\begin{equation}
	\Ra = \frac{\alpha g \Delta T d^3}{\kappa \nu},
\end{equation}
the Ekman number,
\begin{equation}
	\Ek = \frac{\nu}{2\Omega d^2},
\end{equation}
and the Prandtl number,
\begin{equation}
	\Pran = \frac{\nu}{\kappa}.
\end{equation}

We assume that all variables are periodic in the horizontal directions. In the vertical direction, 
the upper and lower boundaries are taken to be perfect thermal conductors, impermeable and stress-free, \ie
\begin{eqnarray}
 	&& \theta = 0 \quad  \mbox{at } z=0,1;
	\label{eq:BCtheta}
	\\
	&& \frac{\partial u_x}{\partial z} =  \frac{\partial u_y}{\partial z} = u_z =0 \quad  \mbox{at } z=0,1 .
	\label{eq:BCu}
\end{eqnarray}
By choosing stress-free, rather than no-slip, boundary conditions, we provide the most advantageous 
conditions for the development of horizontal flows of large amplitude.

Equations~(\ref{eq:u})--(\ref{eq:theta}) are solved numerically using a parallel pseudospectral 
code developed by \citet{Catt03}. The temperature perturbation and each component of the velocity are 
transformed from configuration space (containing \mbox{$N_x \times N_y \times N_z$} collocation points) 
to phase space (containing \mbox{$n_x \times n_y \times n_y$} modes) by a discrete Fourier transform of the form
\begin{equation}
 f(x,y,z) =  \sum\limits_{n_x}  \sum\limits_{n_y}  \sum\limits_{n_z} \hat{f} (n_x,n_y,n_z)
	      \exp (2\pi i k_x x) \exp (2\pi i k_y y) \phi^q (\pi k_z z) + \mathrm{c.c.}, 
\end{equation}
where $f$ and $\hat{f}$ are the functions in configuration and phase spaces respectively, 
$\mathrm{c.c.}$ denotes complex conjugate, $q = \pm 1$ depending on the boundary conditions 
for $f$, with $\phi^{+1}(s) = \cos(s)$ and $\phi^{-1}(s) = \sin(s)$, and
\begin{equation}
 k_x = \frac{n_x}{\lambda}, \quad k_y = \frac{n_y}{\lambda}, \quad k_z = n_z.
\end{equation}
Further details concerning the numerical methods can be found in \citet{Catt03}.

\section{Large-Scale Vortices}
\label{sec:LSV}
In this section, we present the results from our simulations of rotating Boussinesq convection, 
discussing in detail the structure and possible formation mechanism for large-scale vortices. The simulations 
are grouped together as series for which the Ekman number and aspect ratio are fixed; in a given series, the Rayleigh 
number and numerical resolution are varied. The parameter values of the different series are summarised in 
table~\ref{tab:series}. The Prandtl number is set to unity for all of the numerical simulations. The minimum and 
maximum resolutions of each series (corresponding to the smallest and largest $\Ra$ respectively) are also 
given in table~\ref{tab:series}.
The horizontal grid resolution is determined  by the width of the convective structures. 
To ensure a true representation of the flow, we always check that the tail of the kinetic energy spectrum at 
each depth is at least a factor $10^4$ smaller than the peak (as can be observed, for instance, 
in figure~\ref{fig:KE_lambda}) 
and that the Kolmogorov microscale is larger than the minimum grid space.
Since we are imposing stress-free boundary conditions, the vertical grid resolution 
is essentially limited by the thickness of the top and bottom thermal boundary layers; these always contain at 
least 10 collocation points in the vertical. Finally, the timestep is mainly limited by the rotation period. 
For the smallest Ekman number considered here ($\Ek=5\times10^{-6}$) at the largest Rayleigh number 
calculated (see table~\ref{tab:series}), the timestep is $2\times10^{-3}$ (in units of $1/(2\Omega)$).

\begin{table}
\centering
\begin{tabular}{c  c  c  c  c  c c}
\hline
Series & $\Ek$ & $\lambda$ & min($\tRa$) & max($\tRa$) & min resolution & max resolution
\\
 S1 & $10^{-4}$ & $1$ & $10$ & $186$ & $64\times 64\times65$ & $256\times256\times257$
\\
 S2 & $10^{-4}$ & $2$ & $10$ & $186$ & $128\times 128\times65$ & $256\times256\times257$
\\
 S3 & $10^{-4}$ & $4$ & $10$ & $186$ & $256\times 256\times65$ & $512\times512\times257$
\\
 S4 & $10^{-5}$ & $1$ & $10$ & $215$ & $128\times 128\times97$ & $512\times512\times257$
\\
 S5 & $5\times10^{-6}$ & $1$ & $10$ & $188$ & $256\times 256\times129$ & $256\times256\times257$
 \\ \hline
\end{tabular}
\caption{Summary of parameter values and numerical resolution (\mbox{$N_x \times N_y \times N_z$} collocation points) 
for each series of simulations; $\Pran = 1$ and $\tRa = \Ra \Ek^{4/3}$.}
\label{tab:series}
\end{table}

\subsection{Structure}
\label{sec:structure}

At the linear onset of convection, for $\Pran=1$, the flow takes the form of elongated cells with 
velocity and temperature perturbations of vertical wavenumber $k_z=1$ and, for small $\Ek$, 
horizontal wavenumber scaling as $\Ek^{-1/3}$ \citep[e.g.][]{Cha61}. The critical Rayleigh number 
scales as $\Ek^{-4/3}$. Hereafter, in order to compare different sets of simulations, we use the Rayleigh 
number compensated by its Ekman number dependence at the onset of convection, \mbox{$\tRa = \Ra \Ek^{4/3}$}. 
Note that at the onset of convection, $\tRa \approx 8.7$ as $\Ek\to 0$.

\begin{figure}
\centering
  \includegraphics[clip=true,height=4.5cm]{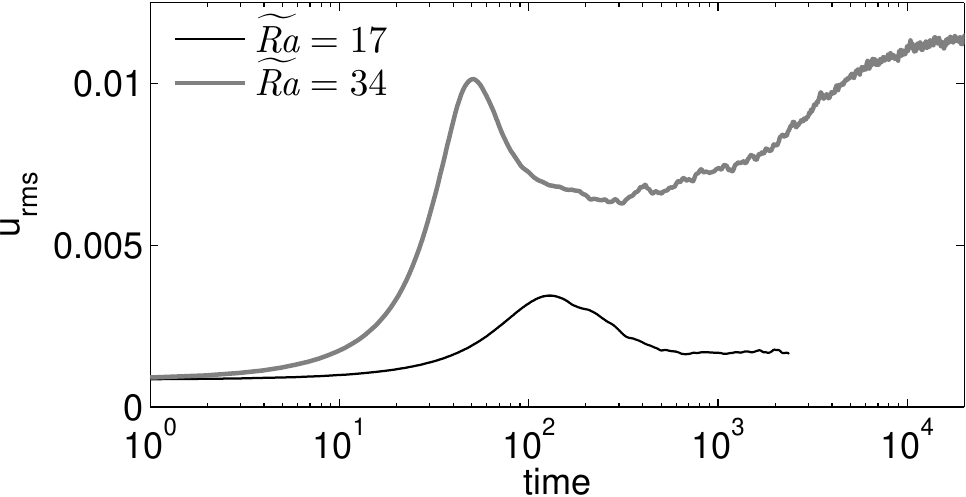}
  \caption{Time series of the rms velocity for $\tRa=17$ and $\tRa=34$ with $\Ek=5\times 10^{-6}$ and $\aspect=1$ (series~S5). 
Since the velocity is scaled by $2\Omega d$  
then the rms velocity is comparable with a Rossby number. 
}
  \label{fig:KE_vs_time}
\end{figure}

Figure~\ref{fig:KE_vs_time} shows time series of the rms velocity (the square root of the kinetic energy 
per unit volume) for the smallest value of the Ekman number we have considered, namely \mbox{$\Ek=5\times 10^{-6}$} 
(series~S5), for $\tRa=17$ and $\tRa=34$. In both cases, the kinetic energy first grows exponentially as 
the convective instability is triggered. For $\tRa=17$, the kinetic energy reaches a stationary state for 
times $t \gtrsim 10^3$ (in units of  $1/(2\Omega)$). However, for $\tRa=34$, after a short period of stagnation, 
the kinetic energy displays a secondary phase of growth at a slower rate; it eventually saturates on a much 
longer timescale than that of the initial convective instability, after a time $t\approx 2\times10^4$, \ie about 
one tenth of a global viscous timescale. It is this slow growth of the kinetic energy that corresponds to the 
formation of a large-scale vortex. 

\begin{figure}
\centering
  \subfigure[]{\label{fig:wz_slice_hcc}
  \includegraphics[clip=true,height=3.6cm]{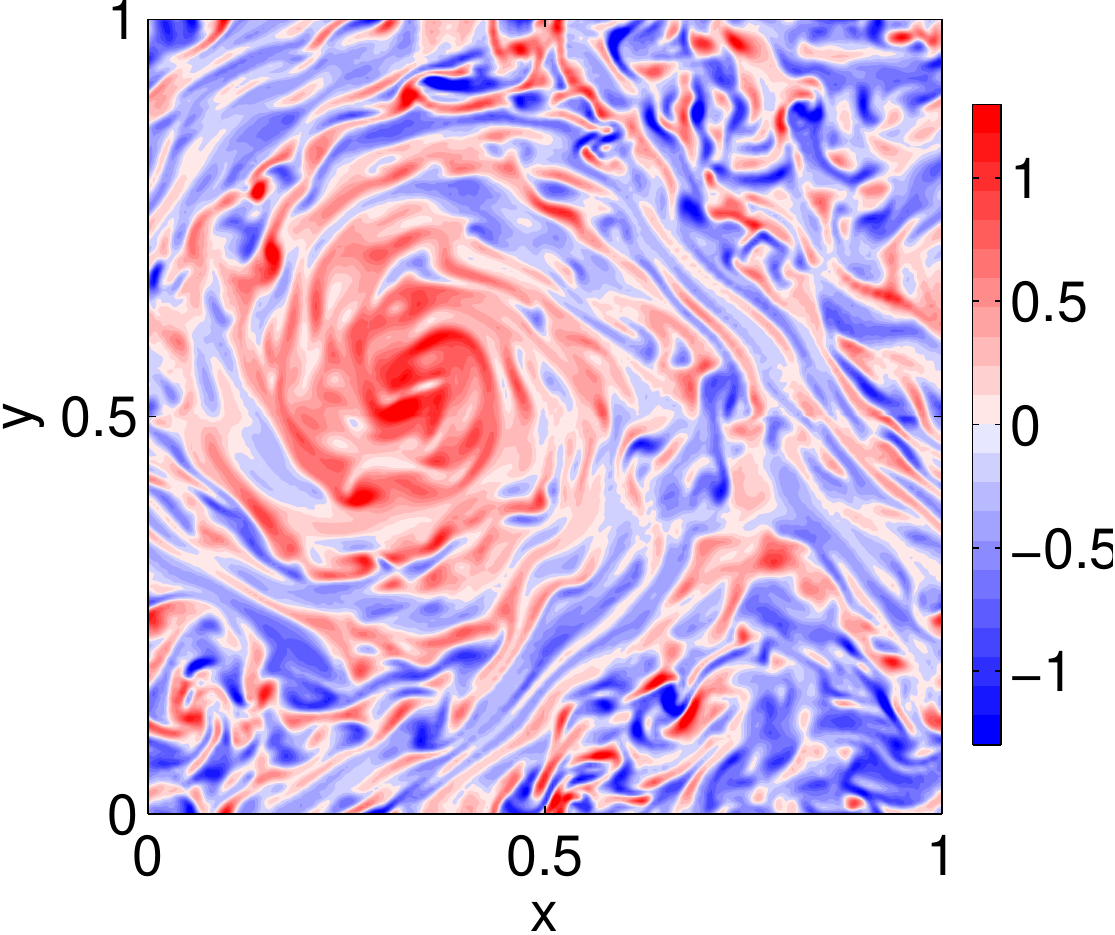}}
  \subfigure[]{\label{fig:wz_slice_vcc}
  \includegraphics[clip=true,height=3.6cm]{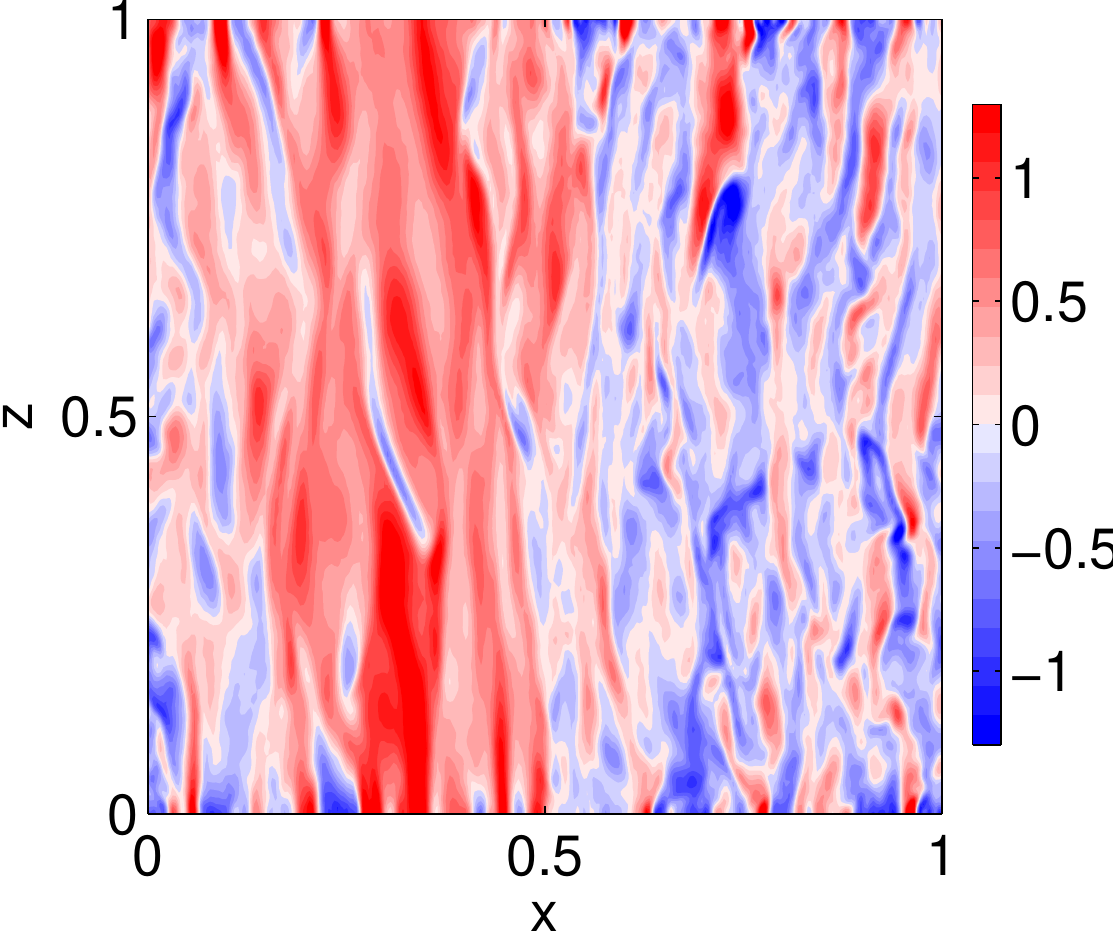}}
  \subfigure[]{\label{fig:wz_ratio}
  \includegraphics[clip=true,height=3.6cm]{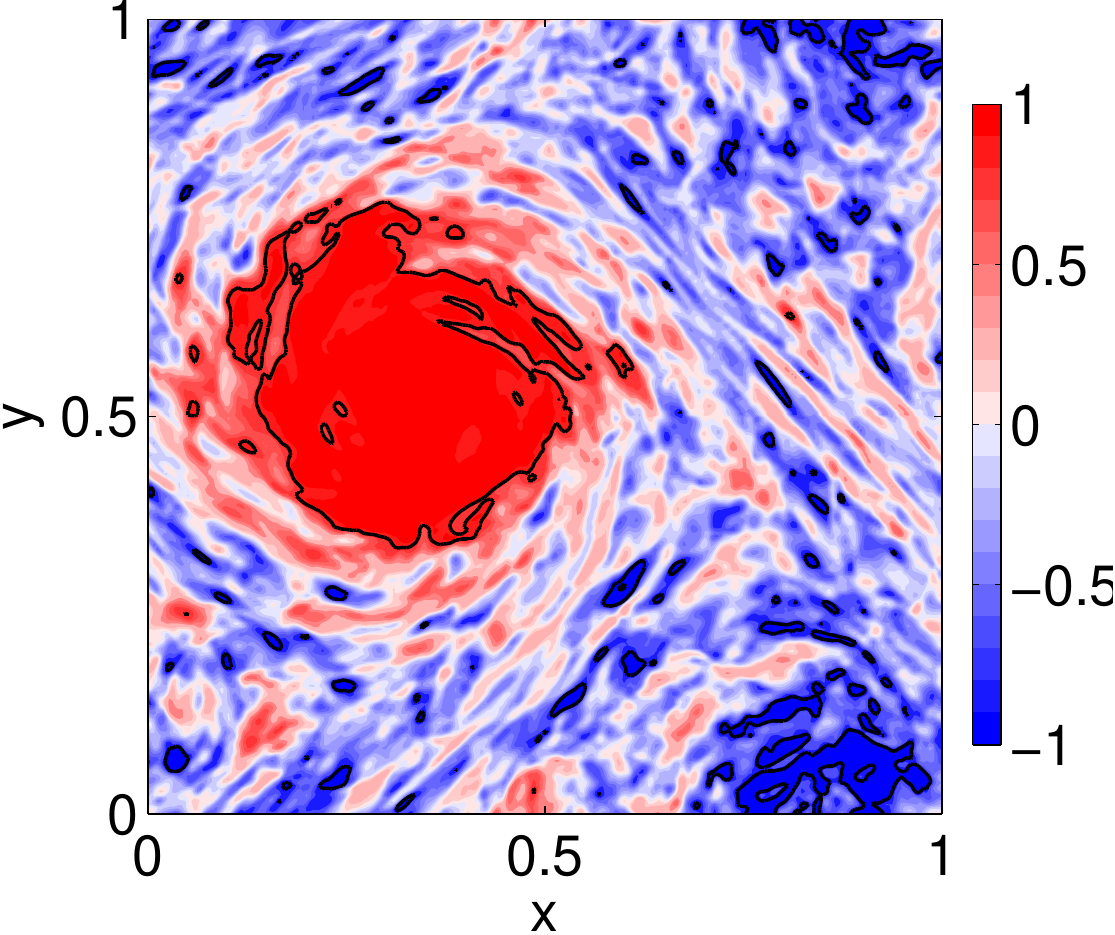}}
  \caption{($a$) Horizontal and ($b$) vertical cross-sections of the axial vorticity
  (snapshot) given in units of $2 \Omega$. 
  The horizontal section is taken at $z=0.25$ and the vertical section at $y=0.50$.
  ($c$) Ratio $r$ measuring the degree of $z$-invariance of the axial vorticity.
  The black lines correspond to the isocontour $\pm 0.8$.
  Parameters: $\tRa=68$ of series~S5.
  \label{fig:wz_slice}}
\end{figure}

Snapshots of the axial vorticity, $\omega_z = (\bnabla \times \vel)\cdot \vect{e}_z$, in horizontal and 
vertical cross-sections during the saturated phase are plotted in figures~\ref{fig:wz_slice_hcc}-\ref{fig:wz_slice_vcc} 
for $\tRa=68$ of series~S5. The flow is organised principally in a cyclonic structure of large horizontal scale,  
surrounded by a multitude of smaller vortices of either sign. The small-scale vortices are driven directly by buoyancy; 
we shall refer to these as \textit{convective structures}. As a consequence of the periodic boundary conditions, 
horizontal averages of the axial vorticity vanish; thus, in a horizontal cross-section, within the multitude of 
small vortices the anticyclonic vorticity must dominate, so as to balance the large-scale cyclonic vorticity. 
Two movies showing the evolution of the axial vorticity in a horizontal cross-section are provided in the online 
supplementary material available at \url{http://dx.doi.org/10.1017/jfm.2014.542}. 
Movie~1 is taken during the slowly growing phase of the kinetic energy, movie~2 during 
the long-term saturated phase.
They demonstrate that the multitude of small vortices are advected by a relatively slow anticyclonic circulation, 
which occupies a larger area than the faster cyclonic circulation. These large-scale circulation cells create 
regions of intense shear, in which small vortices become stretched horizontally. 
For the Rayleigh number of figure~\ref{fig:wz_slice}, the axial vorticity in the core of the large-scale cyclone 
can locally attain an amplitude of the order of the planetary vorticity, $2\Omega$.

The vertical cross-section in figure~\ref{fig:wz_slice_vcc} shows that the large-scale cyclone is predominantly 
$z$-invariant. The degree of $z$-invariance of the axial vorticity can be quantified by the ratio
\begin{equation}
 r(x,y) = \frac{\int \omega_z(x,y,z) dz}{\int ((\omega_z(x,y,z))^2)^{1/2} dz} .
\end{equation}
Figure~\ref{fig:wz_ratio} shows $r(x,y)$ corresponding to the snapshot of 
figures~\ref{fig:wz_slice_hcc}-\ref{fig:wz_slice_vcc}. Inside the core of the large-scale cyclone, 
$r$ is fairly close to unity, implying that the axial vorticity is predominantly $z$-invariant. 
For instance, the isocontour $r=0.8$ corresponds roughly to a circle of diameter $0.3$--$0.4$. 
For the anticyclonic region, on the other hand, $|r|$ is smaller on average, and the regions for 
which $|r|\geq 0.8$ consist essentially of isolated vortex cores of small horizontal extent. 

We were not able 
to detect any secondary circulation associated with the large-scale cyclone. Any such circulation, if present, 
is significantly weaker than the convective motions.

It is important to note that horizontal shear flows (which project on vorticity modes $(k_x, k_y) =(0, 1)$ or $(1, 0)$, 
while the LSV project on mode $(1, 1)$) are permitted in the numerical code; these however were never observed in 
our simulations. The flow is isotropic in the horizontal directions, and there appears to be no mechanism capable 
of driving shear flows consistently in a preferred horizontal direction (as does the $\beta$ effect in the presence 
of a gradient of planetary vorticity, for instance). Indeed, even when a horizontal shear flow of sinusoidal profile 
is added to the convective flow before the formation of LSV and the flow is then allowed to evolve freely, we observe that 
the horizontal shear flow disappears while a large-scale vortex grows.

\subsection{Domain of existence}
\label{sec:domain}
In this subsection, we determine the domain of existence of large-scale vortices in terms of input and output 
parameters based on the rms velocity (which includes all three components) and the rms vertical velocity, which we 
use as an estimate of the typical convective velocity. As discussed above, the LSV consist essentially of horizontal motions. 
Consequently, a comparison of the amplitudes of horizontal and vertical flows is instructive in determining the domain 
of their existence. We define the Rossby number, $\Ro$, and the vertical Rossby number, $\Ro_z$, by
\begin{equation}
 \Ro =  \frac{\langle u_x^2+u_y^2+u_z^2 \rangle^{1/2}}{2 \Omega d} \quad \textrm{and} \quad 
 \Ro_z = \frac{\langle u_z^2 \rangle^{1/2}}{2\Omega d} ,
\end{equation}
where the angle brackets denote a spatial and temporal average. Note that these definitions use the box depth, 
$d$, as the lengthscale.

\begin{figure}
\centering
  \subfigure[]{\label{fig:Roz_Rat}
  \includegraphics[clip=true,height=4.5cm]{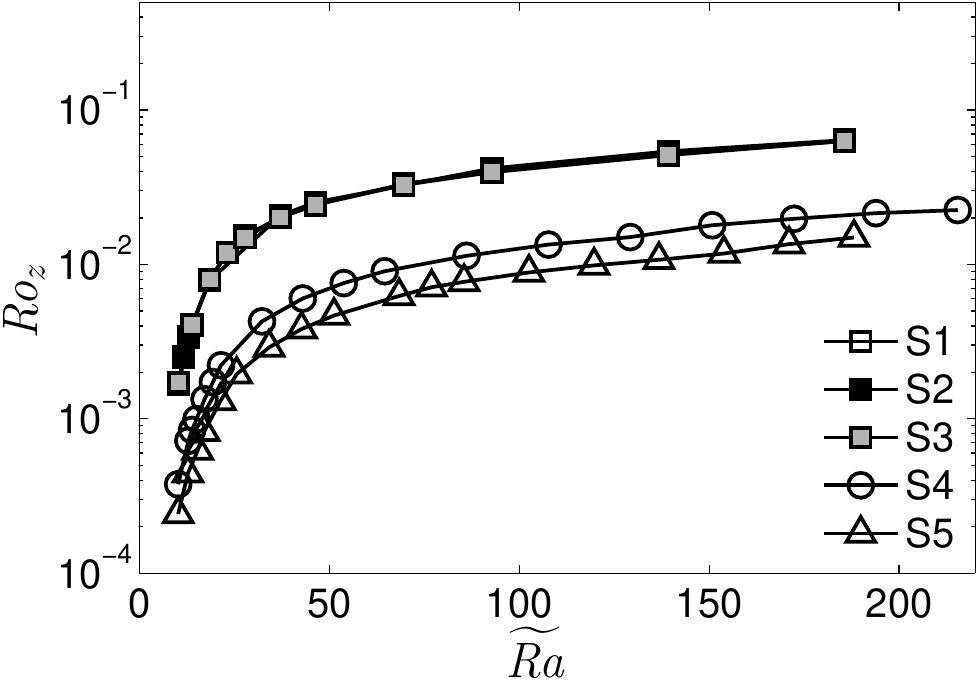}}
  \subfigure[]{\label{fig:Ro_Rat}
  \includegraphics[clip=true,height=4.5cm]{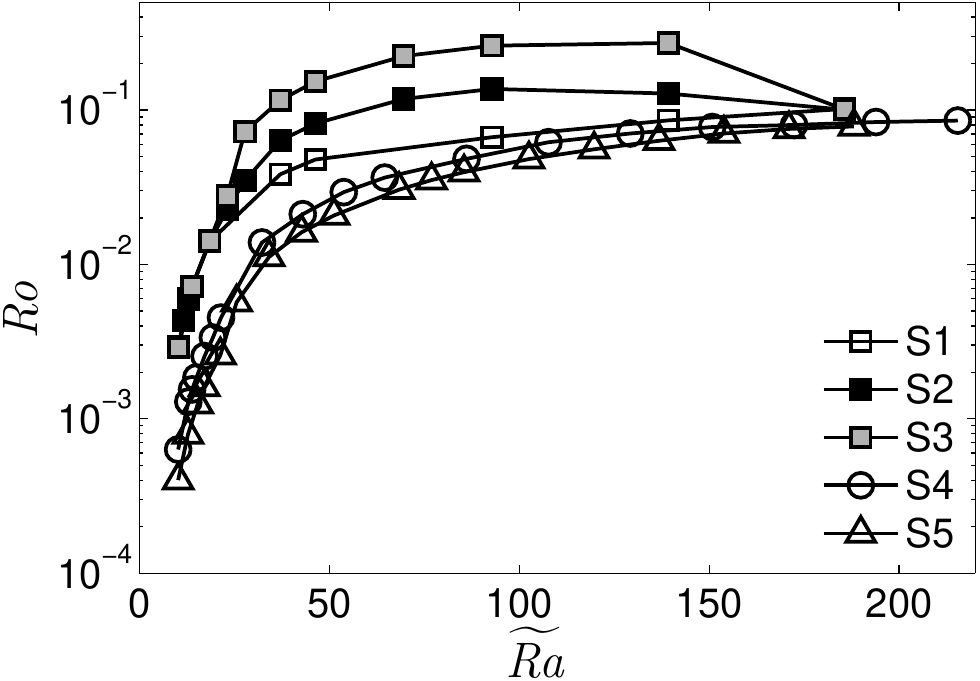}}
   \caption{($a$) Vertical Rossby number, $\Ro_z$; ($b$) Rossby number, $\Ro$, 
as a function of \mbox{$\widetilde{\Ra}$} for the series S1--S5 defined in table~\ref{tab:series}.}
\label{fig:Ro_Roz_Ra}
\end{figure}

Figures~\ref{fig:Roz_Rat}--\ref{fig:Ro_Rat} plot $\Ro_z$ and $\Ro$ versus $\tRa$ for the series~S1--S5. 
$\Ro_z$ is a monotonically increasing function of $\tRa$, and is always smaller than $0.1$ in our simulations. 
In the series~S1--S3, which have the same Ekman number ($\Ek=10^{-4}$) but different aspect ratio, 
the curves of $\Ro_z$ lie on top of each other; varying the aspect ratio in the range considered here 
therefore does not affect the vertical rms velocity. In figure~\ref{fig:Ro_Rat}, the evolution of $\Ro$ 
with $\tRa$ shows important differences compared with that of $\Ro_z$. First, $\Ro$ decreases noticeably 
when $\tRa \gtrsim 150$ in the series~S2--S3. Second, the values of $\Ro$ are not identical for the series~S1--S3 
for $\tRa \gtrsim 20$, thus demonstrating their aspect ratio dependence, with the largest value 
of $\Ro$ occurring for $\lambda=4$.
Thus the amplitude of the horizontal flows does not follow the evolution of the amplitude of the convective flows.

\begin{figure}
\centering
  \subfigure[]{\label{fig:ReRez_Rat}
  \includegraphics[clip=true,width=6.5cm]{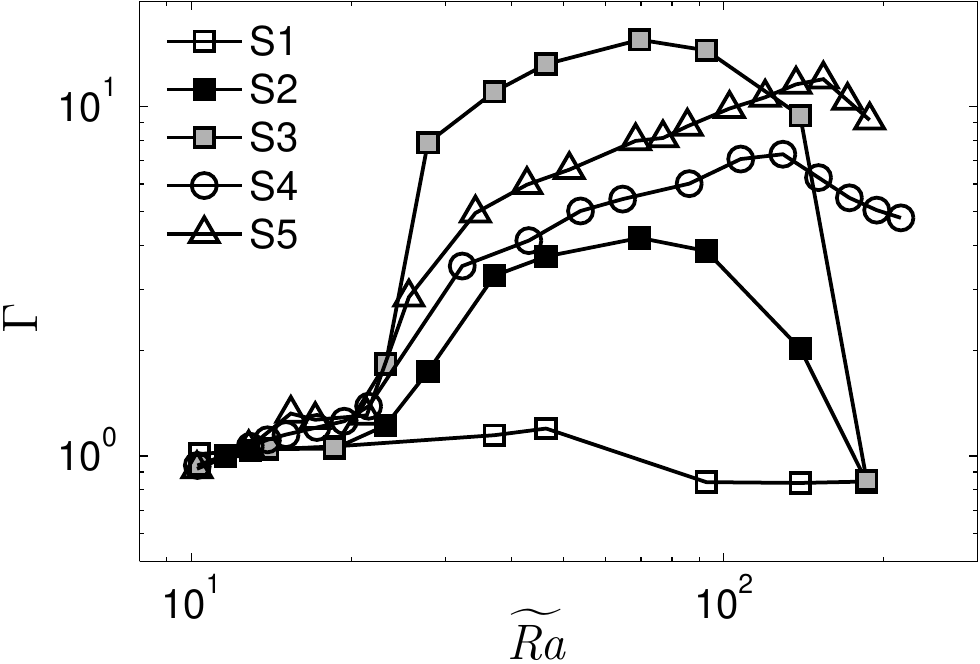}}
  \subfigure[]{\label{fig:ReRez_Rez}
  \includegraphics[clip=true,width=6.5cm]{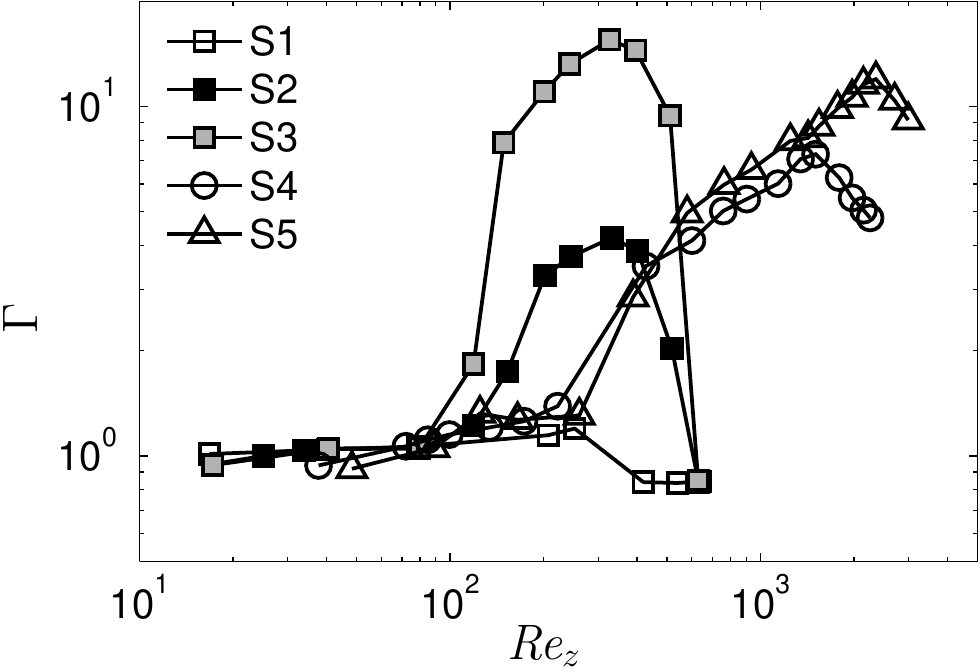}}
  \subfigure[]{\label{fig:ReRez_tilgner}
  \includegraphics[clip=true,width=6.5cm]{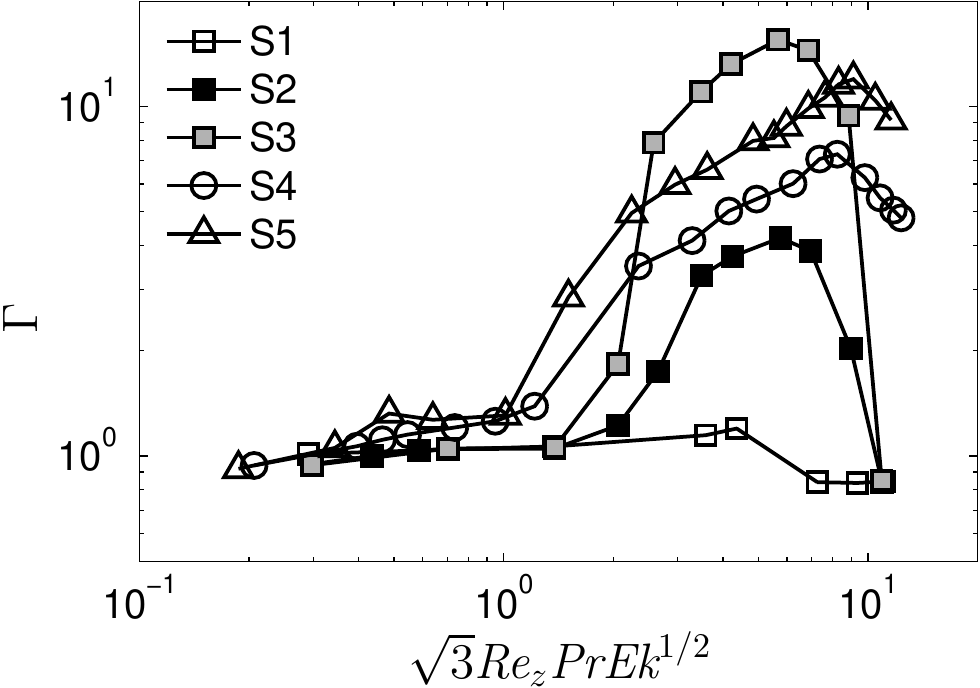}}
  \subfigure[]{\label{fig:ReRez_Rozl}
  \includegraphics[clip=true,width=6.5cm]{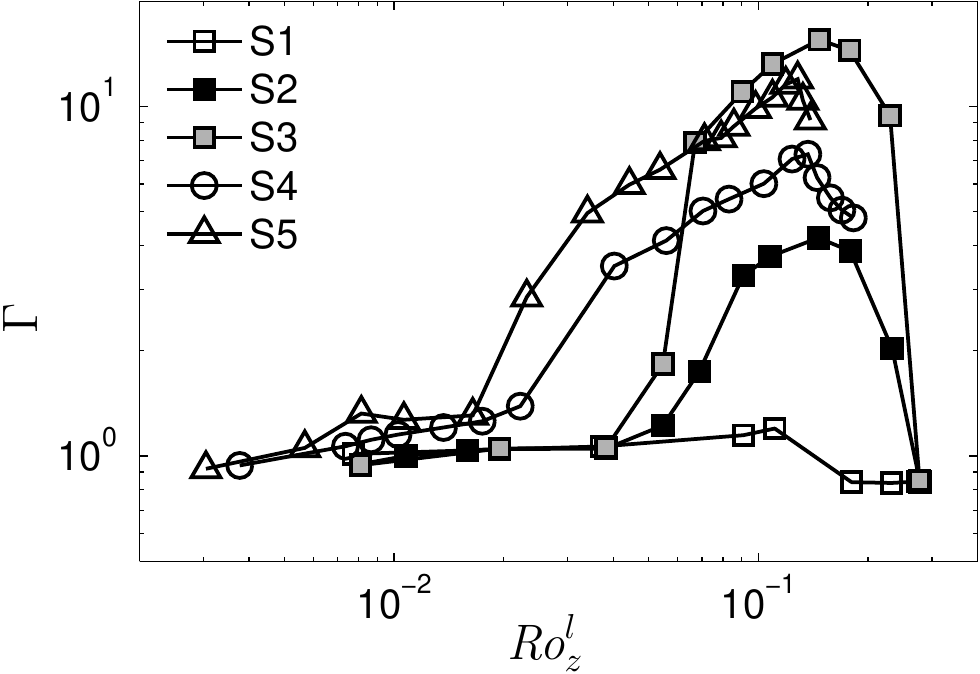}}
  \caption{Ratio of the total kinetic energy to the vertical kinetic energy, \mbox{$\Gamma$},
          as a function of ($a$) \mbox{$\widetilde{\Ra}=\Ra \Ek^{4/3}$}, 
($b$) $\Rey_z$, ($c$) $\sqrt{3}\Rey_z \Pran \Ek^{1/2}$ and ($d$) $\Ro_z^l$ for the series~S1--S5.}
\label{fig:ReRez}
\end{figure}

As a measure of the ratio of the total kinetic energy to the vertical kinetic energy we define
\begin{equation}
 \Gamma = \frac{\langle u_x^2+u_y^2+u_z^2 \rangle}{3 \langle u_z^2 \rangle}.
\label{eq:gamma}
\end{equation}
Figure~\ref{fig:ReRez} shows $\Gamma$ as a function of different input and output parameters
for the series~S1--S5. Figure~\ref{fig:ReRez_Rat} plots $\Gamma$ versus $\tRa$. 
While $\Gamma$ is approximately unity close to the onset of convection, it increases rapidly 
for $\tRa \gtrsim 20$ in the series~S2--S5. For the series~S1, which corresponds to the largest Ekman number 
considered ($\Ek=10^{-4}$) and the smallest aspect ratio ($\aspect=1$), the variations of $\Gamma$ appear 
small on the vertical scale used for the figure. In the series~S2--S5, for large values of $\tRa$ (typically 
larger than $70$ but depending on the Ekman number), $\Gamma$ reaches a maximum, which is greater than $10$ 
in some cases, before eventually decaying. The maximum of $\Gamma$ occurs at increasing values of $\tRa$ for decreasing $\Ek$. 

We now seek the parameters that control the evolution of $\Gamma$. We have already noted that all of the $\Gamma$ curves 
first increase just after $\tRa\approx 20$. In order to quantify the level of turbulence of the convective flow, we measure 
the vertical Reynolds number, defined by
\begin{equation}
  \Rey_z = \frac{\Ro_z}{\Ek} = \frac{\langle u_z^2 \rangle^{1/2} d}{\nu} .
\end{equation}
Figure~\ref{fig:ReRez_Rez} plots $\Gamma$ as a function of $\Rey_z$. The increase of  $\Gamma$ 
is sharp after some value of $\Rey_z$, $\Rey_z^*$ say, that is dependent on $\lambda$ but not on $\Ek$. 
For $\lambda=1$, $\Rey_z^* \approx 300$ and becomes smaller for larger aspect ratios; for instance, 
when $\lambda=4$, $\Rey_z^* \approx 100$. The existence of a threshold vertical Reynolds number for the appearance of 
the LSV implies that a certain level of convectively driven turbulence is required. However, $\Rey_z^*$ takes rather 
modest values, achieved for Rayleigh numbers only about three times its critical value at the onset of convection.

The maximum in $\Gamma$ occurs at increasing values of $\Rey_z$ for decreasing $\Ek$, indicating that the decrease 
of $\Gamma$ could be due to a transition from a convection regime that is strongly rotationally constrained, to one 
that is only weakly constrained. Using a similar numerical model of rotating Boussinesq convection, \citet{SchTil09} 
find empirically, through measurements of the heat flux, that the transition from rapidly rotating convection to 
weakly rotating convection occurs when $\Rey \Pran \Ek^{1/2} \approx 10$, where $\Rey$ is a Reynolds number based 
on the rms velocity and the box height. \citeauthor{SchTil09} do not mention the presence of LSV in their simulations, 
so we take $\sqrt{3}\Rey_z$ as being equivalent to their Reynolds number. Figure~\ref{fig:ReRez_tilgner} 
shows $\Gamma$ as a function of \mbox{$\sqrt{3} \Rey_z \Pran \Ek^{1/2}$}. It can be seen that the maxima in the 
curves of $\Gamma$ are not strictly aligned at the value $10$, but instead tend to occur at smaller values 
of \mbox{$\sqrt{3} \Rey_z \Pran \Ek^{1/2}$} as $\Ek$ increases. 

To measure the influence of rotation, we use a more traditional dimensionless quantity, 
the local Rossby number, $Ro_z^l$, defined by
\begin{equation}
 \Ro_z^l = \frac{\langle u_z^2 \rangle^{1/2}}{2\Omega l_h} = \frac{\Ro_z}{l_h/d};
\end{equation} 
the dimensionless horizontal lengthscale, $l_h^{\ast}=l_h/d$, is defined by
 \begin{equation}
   {l_h^{\ast}}^{-1} =  \left\langle 
               \frac{ \sum\limits_{k_x,\, k_y,\, k_z}  \sqrt{k_x^2+k_y^2} \pleft \hat{u}_z(k_x, k_y, k_z) \pright^2}
		   {\sum\limits_{k_x,\, k_y,\, k_z} \pleft \hat{u}_z(k_x, k_y, k_z) \pright^2} \right\rangle,
\label{eq:lh}
\end{equation}
where the vertical velocity is expressed in spectral form $\hat{u}_z$. Figure~\ref{fig:ReRez_Rozl} 
shows $\Gamma$ as a function of $\Ro_z^l$. The lengthscale $l_h^{\ast}$ depends on the Ekman number, and scales 
as $\Ek^{1/3}$ close to the onset of convection. 
Since the definition of $l_h^{\ast}$ is based on the vertical velocity, which receives only a small 
contribution from the LSV, $l_h^{\ast}$ is not significantly affected by the LSV. As $\Ra$ increases in a given series, 
the convective structures tend to become wider, \ie $l_h^{\ast}$ increases. However, $\Ro_z$ increases more rapidly 
with $\Ra$ than $l_h^{\ast}$; thus $\Ro_z^l$ is a monotonically increasing function of $\Ra$ in a given series. 
The maximum of $\Gamma$ occurs for a similar value of $\Ro_z^l$, about $0.15$, for the series~S2--S5.

\begin{figure}
\centering
  \subfigure[$\lambda=1$]{\label{fig:wz_Ta1e8Ra8e6_l1}
  \includegraphics[clip=true,width=4.4cm]{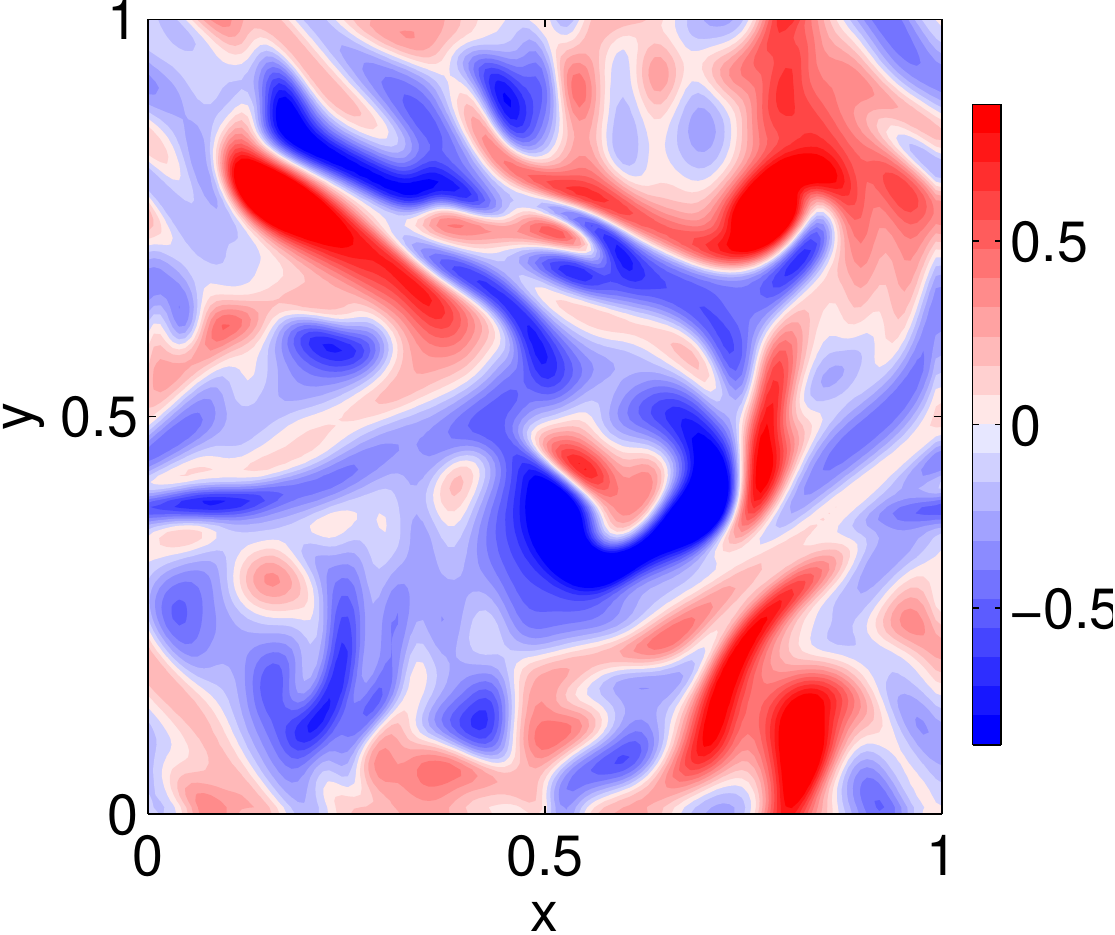}}
  \subfigure[$\lambda=2$]{\label{fig:wz_Ta1e8Ra8e6_l2}
  \includegraphics[clip=true,width=4.2cm]{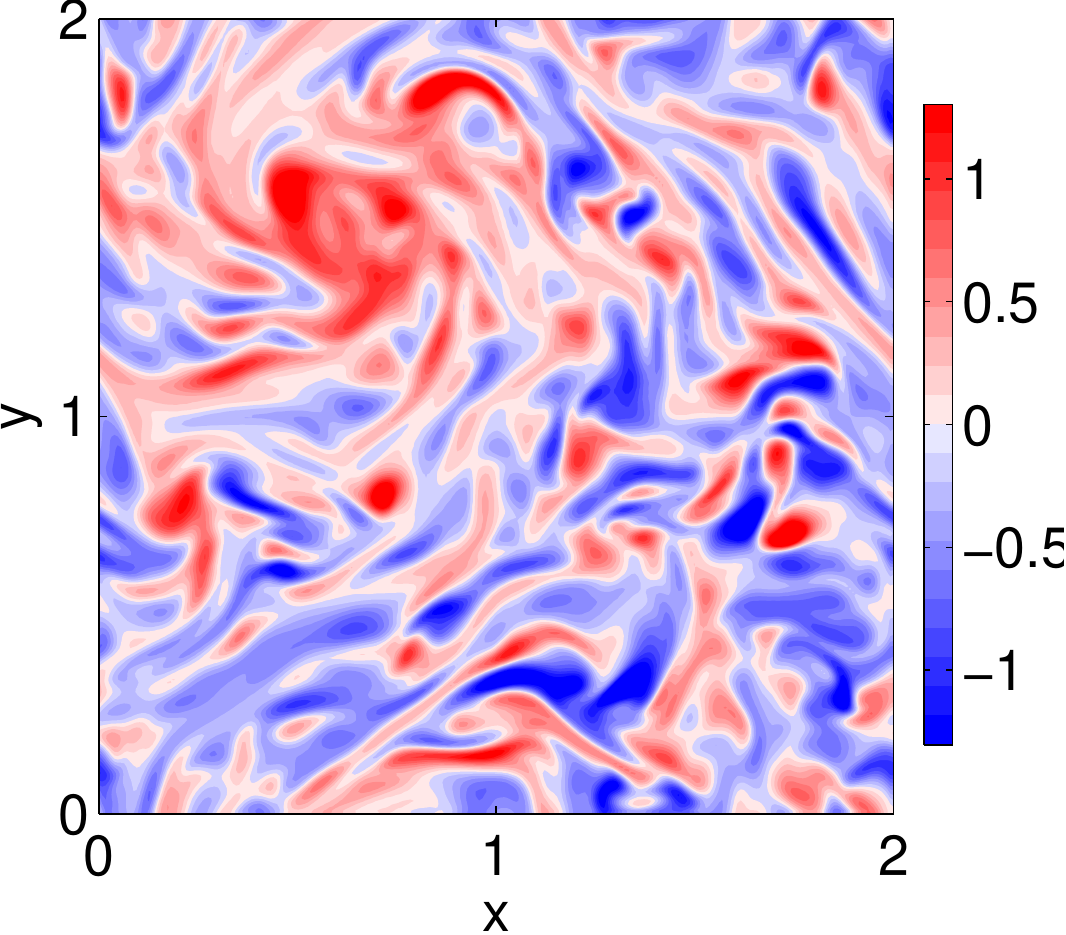}}
  \subfigure[$\lambda=4$]{\label{fig:wz_Ta1e8Ra8e6_l4}
  \includegraphics[clip=true,width=4.2cm]{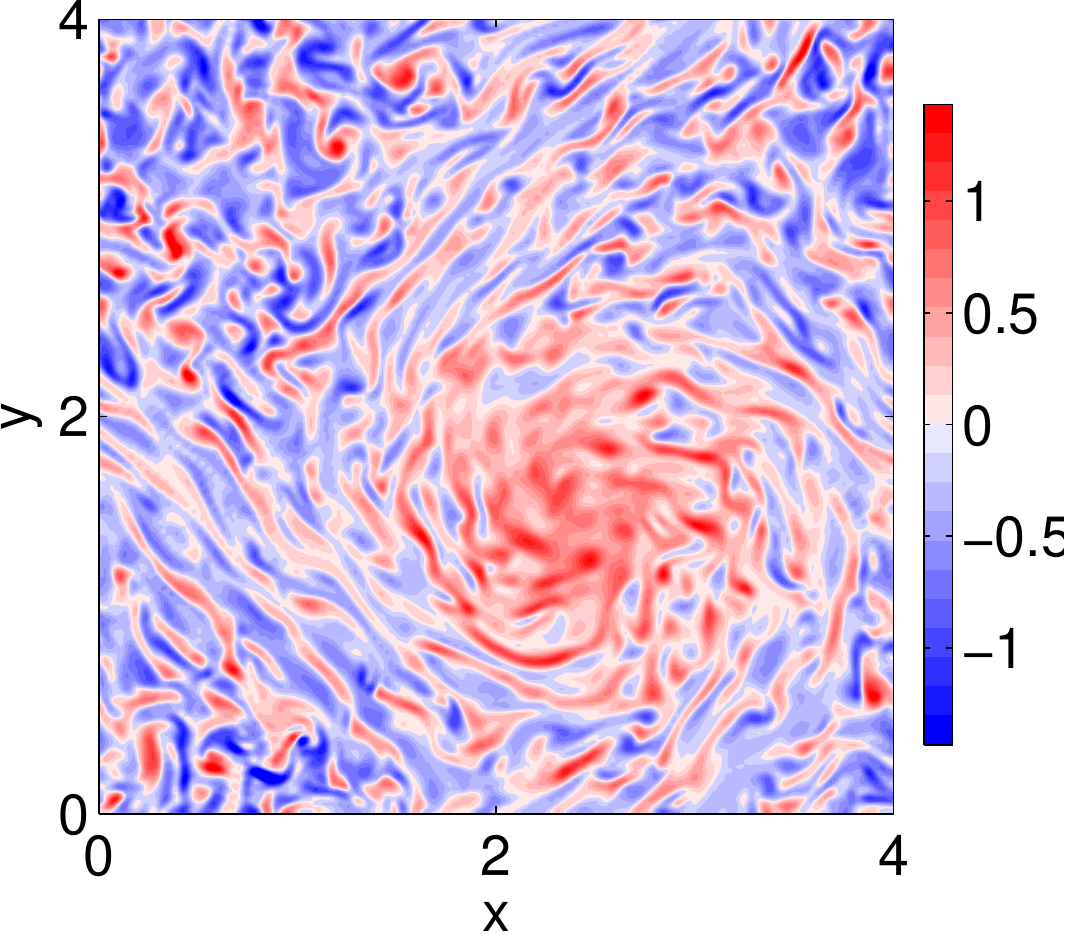}} 
  \caption{Horizontal cross-sections (at $z=0.25$) of the axial vorticity (snapshot) 
for different aspect ratios; $\Ek=10^{-4}$, $\tRa=37$.}
\label{fig:wz_lambda}
\end{figure} 

We now discuss the influence of the aspect ratio $\lambda$ on the existence of LSV in series~S1--S3. 
Figures~\ref{fig:Ro_Roz_Ra} and~\ref{fig:ReRez} show that outside the parameter window in which LSV 
occur (i.e.\ when $\tRa<20$ and $\Ro_z^l>0.15$), the amplitude of the horizontal and vertical flows is essentially 
independent of the aspect ratio, because several convective structures can be accommodated horizontally even for 
the series~S1, which has the smallest aspect ratio. However, in the regime in which LSV are present,
figure~\ref{fig:ReRez} shows clearly that the aspect ratio influences the amplitude of the horizontal flows. 
In the series~S1, values of $\Gamma$ remain close to unity compared with the series~S2--S3. Nonetheless, the evolution 
of $\Gamma$ with $\tRa$ for the series~S1 follows the same trend as the other series, reaching a maximum of $1.2$ 
for $\tRa=46$. Figure~\ref{fig:wz_lambda} shows horizontal cross-sections of the axial vorticity for the series~S1--S3 
at $\tRa=37$. For $\lambda=2$ and $\lambda=4$, cyclonic circulation is visible at a large horizontal scale.  
By contrast, visual inspection of the axial vorticity for $\lambda=1$ 
does not reveal the presence of a large-scale vortex since the scales of the convective structures 
are not much smaller than the box size. A cross-section of the vorticity tends to emphasise smaller scales than that of 
the velocity, so we examine instead the kinetic energy spectra. For a given horizontal wavenumber, 
$k_h=(k_x^2+k_y^2)^{1/2}$, we define the energy spectrum of the horizontal velocity, $\vect{u}_h=(u_x,u_y,0)$, by
\begin{eqnarray}
  E_h (k_h) = \frac{1}{2} \sum\limits_{k_z} \sum\limits_{k_x,k_y} \hat{\vect{u}}_h (k_x,k_y,k_z) \cdot 
	\hat{\vect{u}}_h^{\ast} (k_x,k_y,k_z) ,
	\label{eq:Eh}
\end{eqnarray}
and the energy spectrum of the vertical velocity, $(0,0,u_z)$, by
\begin{eqnarray}
  E_v (k_h) = \frac{1}{2} \sum\limits_{k_z} \sum\limits_{k_x,k_y} \hat{u}_z (k_x,k_y,k_z) \cdot 
	\hat{u}_z^{\ast} (k_x,k_y,k_z) ,
	\label{eq:Ev}
\end{eqnarray}
where $\ast$ denotes the complex conjugate. The kinetic energy spectra are obtained by binning into rings of 
radius $k_h$ with $\Delta k_h = 1/\lambda$.

\begin{figure}
\centering
  \subfigure[horizontal flow]{\label{fig:spectrum_hor_lambda}
  \includegraphics[clip=true,width=5.5cm]{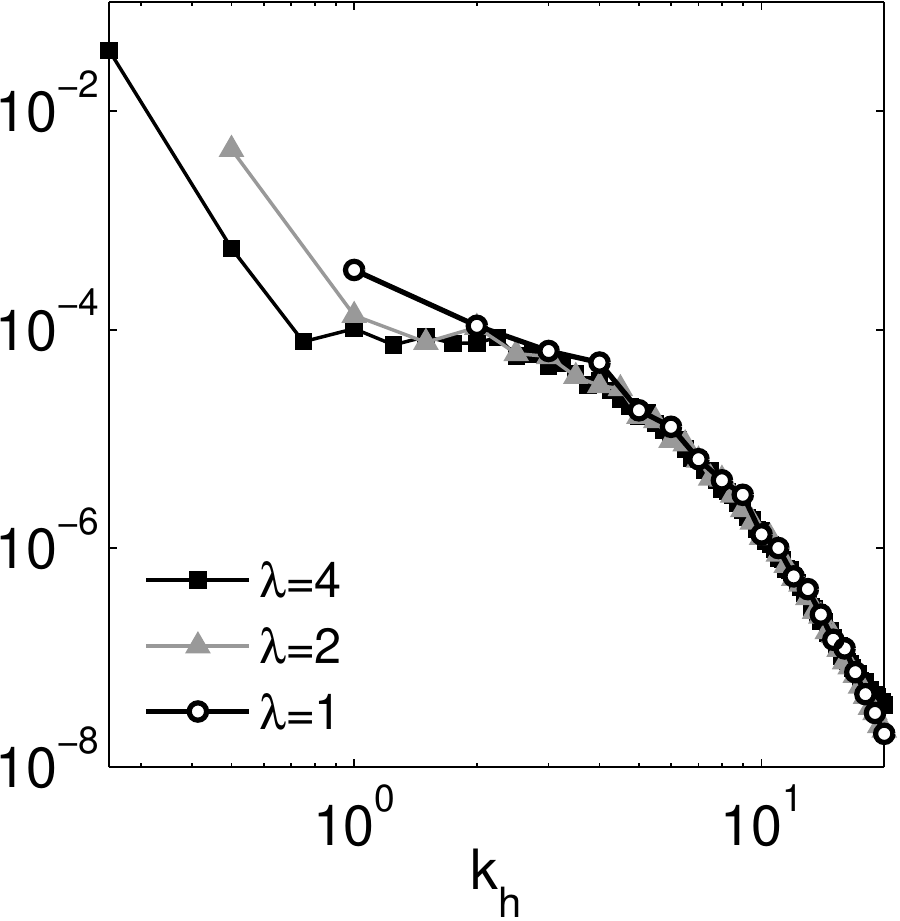}}
  \subfigure[vertical flow]{\label{fig:spectrum_vert_lambda}
  \includegraphics[clip=true,width=5.5cm]{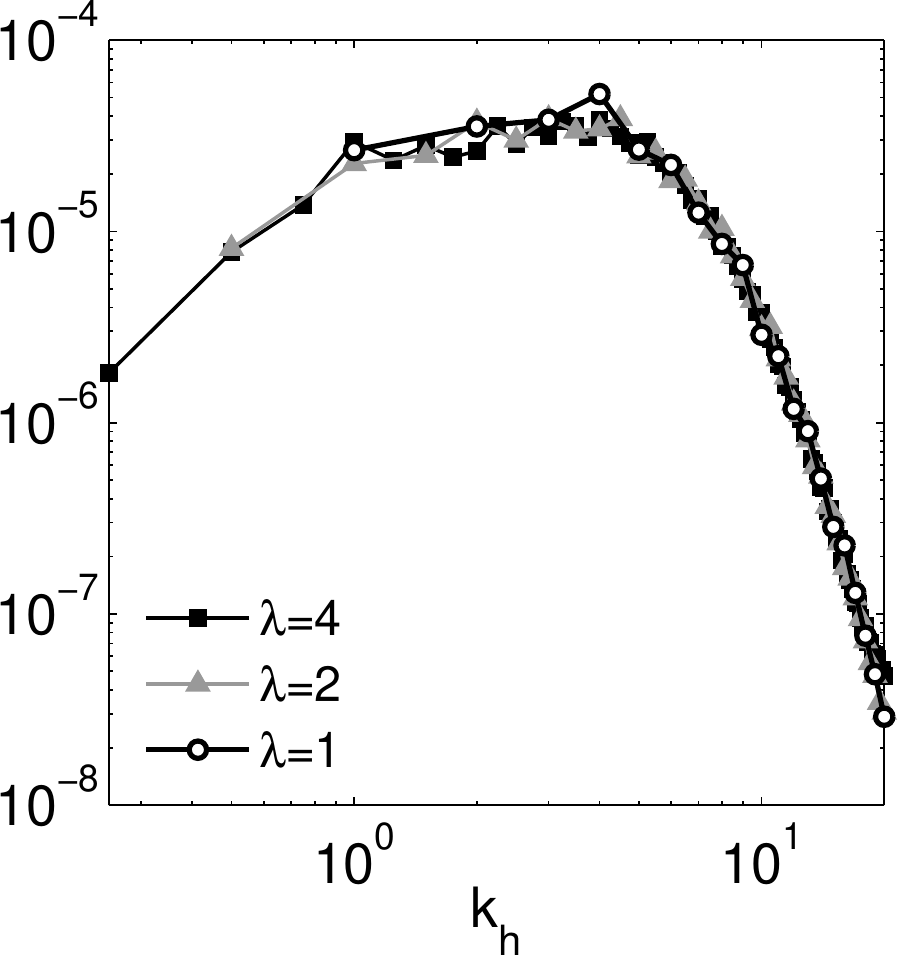}}
  \caption{Kinetic energy spectra (time averages) of the horizontal and vertical flows in the horizontal directions, 
with $k_h=(k_x^2+k_y^2)^{1/2}$. Same parameters as in figure~\ref{fig:wz_lambda}.}
\label{fig:KE_lambda}
\end{figure} 

Figure~\ref{fig:KE_lambda} shows the time-averaged kinetic energy spectra $E_h(k_h)$ and $E_v (k_h)$ for $\tRa=37$ of 
series~S1--S3. They show that the horizontal flow is dominated by the smallest permitted horizontal wavenumber 
for $\lambda=1$, $2$ and $4$, whereas the vertical flow is dominated by the horizontal wavenumber of the convective
structure, \ie $k_h\approx 4$ for $\Ek=10^{-4}$. As $\lambda$ increases, the amplitude of the horizontal flow at the smallest 
horizontal wavenumber becomes larger. As mentioned above, the saturation of kinetic energy occurs on a long timescale, 
of the order of one tenth of a global viscous timescale. This suggests that the saturation of the LSV occurs once a balance 
is established between the viscous dissipation of the LSV and the nonlinear interactions feeding it. Here, the use of 
stress-free boundary conditions ensures that viscous damping of the LSV occurs preferentially in the bulk of the fluid 
rather than in the boundary layers, unlike what would be expected for no-slip boundaries. Consequently, as the LSV increase 
in size with increasing $\lambda$, the horizontal flows can grow to larger amplitudes before being dissipated viscously.

In summary, we find that there are two conditions for the presence of LSV. 
(i)~The Reynolds number of the vertical flows ($\Rey_z$) must be larger than about $100-300$, 
depending on the aspect ratio; this value of the vertical Reynolds number is reached for Rayleigh numbers 
about three times that at the onset of convection. 
(ii)~The convection remains in a regime strongly dominated by rotation, where the local vertical Rossby 
number ($\Ro_z^l$) is smaller than about $0.15$. This value of $\Ro_z^l$ seems robust to changes in the Ekman number 
and the aspect ratio. 
For all of the series, this value of $\Ro_z^l$ corresponds to a similar degree of anisotropy of 
the convective structures, which we measure by $l_h^{\ast}/l_z^{\ast}$; here $l_h^{\ast}$ is given by equation~(\ref{eq:lh}), 
and $l_z^{\ast}$ is defined in a similar manner with $k_z/2$ replacing $k_h$ in the numerator. When $\Ro_z^l$ increases, 
$l_h^{\ast}/l_z^{\ast}$ also increases, and we find that $\Ro_z^l=0.15$ corresponds to $l_h^{\ast}/l_z^{\ast}\approx1/4$,  
so the convective structures must retain a significant degree of anisotropy. Moreover, if conditions (i) and (ii) are 
satisfied, then even in the case of a modest scale separation between the horizontal extent of the convective 
structures ($l_h^{\ast}$) and the horizontal box size ($\lambda$) ($\lambda/l_h^{\ast} \approx 4$ is the smallest scale 
separation considered), energy transfer from the convective size to the large scale
still takes place. This can be identified in kinetic energy spectra of the horizontal flow, which peak at the smallest 
wavenumber, even though in this case LSV are not readily apparent in a visual inspection of the axial vorticity.

\subsection{Asymmetry between cyclones and anticyclones}
\label{sec:cyclonic}

In all of our simulations that produce large-scale structures, 
visual inspection reveals
a concentrated patch of cyclonic vorticity situated in a sea of predominantly anticyclonic 
vorticity (\eg figure~\ref{fig:wz_slice_hcc}). The large cyclonic vortex is stable in time, in the sense that its sign 
does not change and its axial vorticity undergoes only small fluctuations of amplitude compared with its mean value. 
By contrast, in the compressible convection simulations of \citet{Kap11} and \citet{Chan13}, large-scale concentrated 
anticyclones appear at small Rossby numbers, while cyclones are obtained for larger Rossby numbers (although still 
smaller than unity). (Note that Rossby numbers are defined differently in the compressible and Boussinesq cases, 
so the values are not directly comparable.)

In this section, we first establish systematically that the distribution of $\omega_z$ is statistically 
skewed towards cyclonic vorticity in the presence of LSV. We then examine what, in very broad terms, may be regarded 
as the two possible causes of the cyclonic/anticyclonic asymmetry in our system. One is that the nonlinear mechanism 
that transfers energy to the large scales works in favour of cyclonic vorticity. The other is that the generation 
mechanism of large-scale structures favours neither cyclones nor anticyclones, but that any anticyclones formed are 
subsequently unstable.

In a horizontally periodic domain, horizontal averages of $\omega_z$ vanish identically at all depths; a global 
measure of the asymmetry between cyclones and anticyclones is thus provided by the axial vorticity skewness 
\citep[e.g.][]{Bar94}, defined by
\begin{equation}
 S=\frac{\langle \omega_z^3\rangle}{\langle \omega_z^2\rangle^{3/2}} .
\label{eq:S}
\end{equation}
The angle brackets denote both spatial averages, taken over the domain, and temporal averages, 
calculated during the saturated phase of the kinetic energy. 
If $S\ne 0$, the probability density function (p.d.f.) of $\omega_z$ 
is asymmetrical about its mean; a positive (negative) sign of $S$ indicates that the right (left) side of the 
tail of the p.d.f. is either longer or fatter.

\begin{figure}
\centering
  \subfigure[]{\label{fig:S_Ra}
  \includegraphics[clip=true,height=4.5cm]{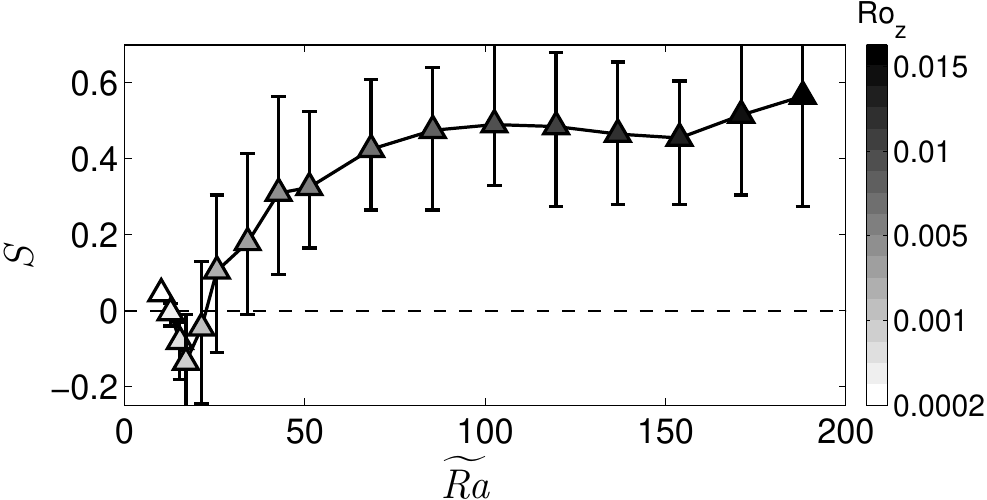}}
  \subfigure[]{\label{fig:Sh_Ra}
  \includegraphics[clip=true,height=4.5cm]{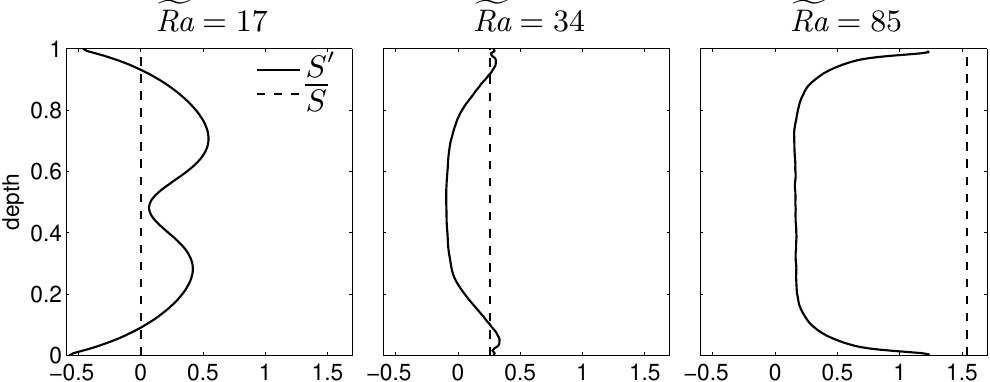}}
    \caption{($a$) Axial vorticity skewness versus $\tRa$ for series S5. 
    The shading of the symbol corresponds to $\Ro_z$, with the values indicated in the grey scale. 
The error bars indicate the minimum and maximum values of the skewness during the time integration 
(in the saturated phase of the kinetic energy).
    ($b$) Vertical profile of the $z$-dependent axial vorticity skewness $S'$ (solid line), and of 
the $z$-invariant axial vorticity skewness $\overline{S}$ (dashed line), for three cases of series~S5.
}
\end{figure}

Figure~\ref{fig:S_Ra} shows $S$ as a function of $\tRa$ for the series~S5. The error bars indicate 
the minimum and maximum values of the skewness during the time integration. The shading of the symbols indicates 
the value of the vertical Rossby number, $\Ro_z$, allowing us to determine whether large-scale anticyclones form 
in our simulations at low $\Ro_z$ but then disappear at higher $\Ro_z$. For $\tRa \gtrsim 20$, $S$ is always positive, 
implying that cyclonic vorticity
is favoured. For $\tRa \lesssim 20$, the values of $S$ are close to zero, with the zero value within the error bar. 
In the cases $15 \lesssim \tRa \lesssim 20$, $S$ takes small negative values. However, for these small Rayleigh numbers, 
LSV are not identifiable either in the velocity field or the kinetic energy spectra. 

Although the skewness $S$ defined by expression~(\ref{eq:S}) is a widely used single measure of asymmetry, 
it is helpful here to look into its constituents in a little more detail. Thus to determine if the sign of $S$ is due mainly 
to contributions from the $z$-invariant axial vorticity, \ie the LSV, or from the $z$-dependent axial vorticity, \ie the 
convective structures, it is instructive to calculate the skewness of the $z$-invariant axial vorticity, 
\begin{equation}
 \overline{S}=\frac{\langle \overline{\omega}_z(x,y)^3 \rangle_h}{\langle \overline{\omega}_z(x,y)^2 \rangle_h^{3/2}} ,
\end{equation}
and the vertical profile of the $z$-dependent axial vorticity skewness,
\begin{equation}
 S'(z)=\frac{\langle \omega'_z(x,y,z)^3 \rangle_h}{\langle \omega'_z(x,y,z)^2 \rangle_h^{3/2}} ,
\label{eq:Sh}
\end{equation}
where $\langle (\cdot) \rangle_h$ denotes time and horizontal averages, and the $z$-dependent
axial vorticity is
\begin{equation}
 \omega'_z(x,y,z) = \omega_z(x,y,z) - \overline{\omega}_z(x,y),
\end{equation}
with $\overline{(\cdot)}$ the vertical average.
The profile of $S'$ compared with that of $\overline{S}$ is shown in figure~\ref{fig:Sh_Ra} 
for $\tRa=17$, $\tRa=34$ and $\tRa=85$ of series~S5. For $\tRa=17$, $\overline{S}\approx 0$, but 
for $\tRa=34$ and $\tRa=85$, $\overline{S}$ takes $O(1)$ positive values, suggesting that positive values of $S$ 
are due in large part to the presence of the large-scale depth-invariant cyclone.

The values of $S'$ in figure~\ref{fig:Sh_Ra} are strongly dependent on $z$ for the three values of $\tRa$. 
For $\tRa=17$, $S'$ is approximately symmetric with respect to $z=1/2$ and is positive above and below the mid-plane 
and negative near the boundaries. The profile of $S'$ is strikingly different for $\tRa=34$ and $\tRa=85$; it is still 
symmetric with respect to $z=1/2$, but with large positive values near the top and bottom boundaries. 
Towards the mid-plane, $S'$ becomes close to zero and even slightly negative for $\tRa=34$. The change in the shape 
of $S'(z)$ roughly coincides with the formation of the LSV , which is around $\tRa \approx 20$ in all the series.

\begin{figure}
\centering
  \subfigure[$\tRa=34$]{
  \includegraphics[clip=true,height=5cm]{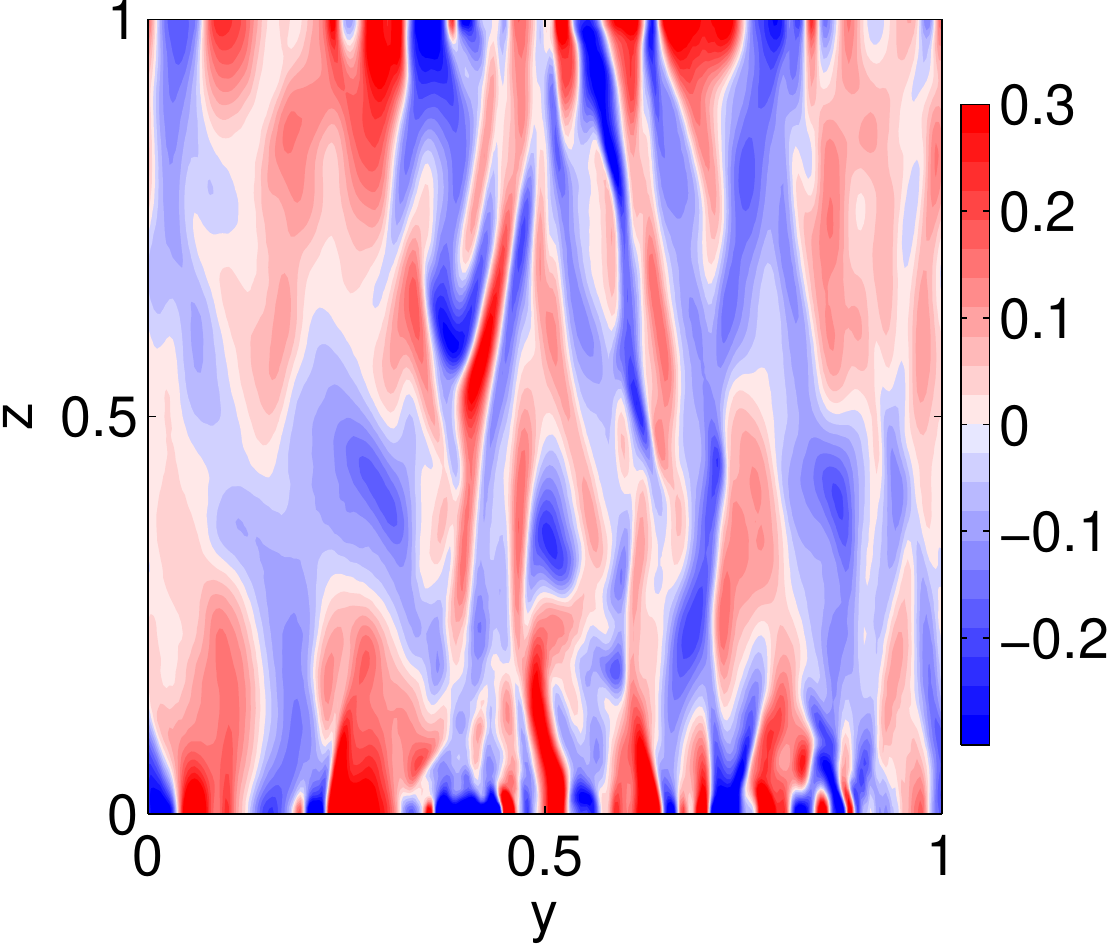}}
  \subfigure[$\tRa=85$]{
  \includegraphics[clip=true,height=5cm]{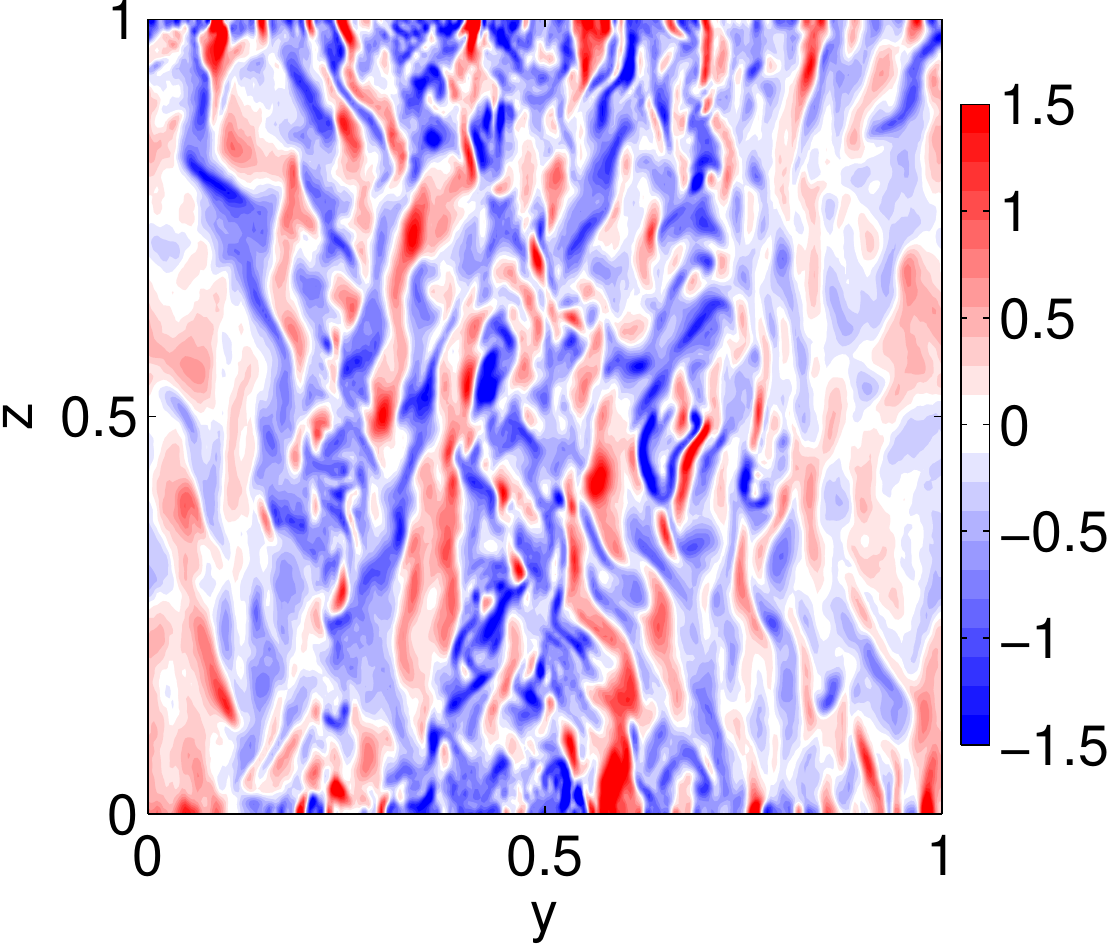}}
    \caption{Vertical cross-sections of the axial vorticity in a $yz$-plane 
    located in the surroundings of the large-scale cyclone for two cases of series~S5.}
\label{fig:wz_plumes}
\end{figure}

Interestingly, this value of $\tRa$ for $\Pran=1$ is identified by \citet{Jul12} as a transition 
in the organisation of the convective structures; for $\tRa \lesssim 20$, the flow consists of cells with a 
high degree of horizontal and vertical coherence, which they denote as cellular convection, whereas for $\tRa \gtrsim 20$, 
thermal plumes develop from a buoyant instability of the thermal boundary layers. In the small Rossby number regime, 
the instability mechanism in the thermal boundary layer permits plume ejection and injection, whereas for the Rossby numbers 
considered here, the mechanism consists solely of plume ejection \citep{Vorobieff2002,Sprague2006}. An important property 
of the thermal plumes in this latter case is that they break the mid-plane symmetry of the axial vorticity 
\citep{Chen1989,Julien1996}. When a thermal plume develops either from the top or bottom thermal boundary layer, it 
drives a convergent horizontal flow by mass conservation, and so acquires cyclonic vorticity by angular momentum conservation; 
as the plume moves toward the opposite boundary, the divergent horizontal flow causes it to spread horizontally and reduces 
its vorticity. Consequently, in the plume regime, the vorticity distribution of the convective flow is expected to be 
skewed toward positive values near the boundaries. 
This is indeed what is seen in figure~\ref{fig:Sh_Ra} for $\tRa=34$ and $\tRa=85$. 
The presence of narrow thermal cyclonic plumes near the top and bottom boundaries can be directly observed 
in figure~\ref{fig:wz_plumes}, which shows vertical cross-sections of the axial vorticity in a $yz$-plane located 
in the surroundings of the large-scale cyclone, taken from snapshots with $\tRa=34$ and $\tRa=85$. On average, 
the vorticity of the plumes changes sign before they reach the mid-layer. The vertical profile of $S'$ indicates 
that the narrow concentrated cyclones extend vertically from the boundaries to a depth of about 20\% of the box height 
for $\tRa=34$. This vertical extension of the cyclonic plumes tends to decrease for larger $\tRa$. The net skewness 
associated with the $z$-dependent flows (\ie when the spatial averages in equation~(\ref{eq:Sh}) are taken over 
the whole domain) is positive, so the positive skewness associated with the cyclonic plumes near the boundaries outweighs 
the negative skewness associated with the anticyclones in the bulk.
Since the vorticity distribution of the convective structures is already skewed towards positive values, it seems plausible 
that a concentrated patch of cyclonic vorticity is more likely to form in the first place.

\begin{figure}
\centering
  \subfigure[]{
  \includegraphics[clip=true,height=4cm]{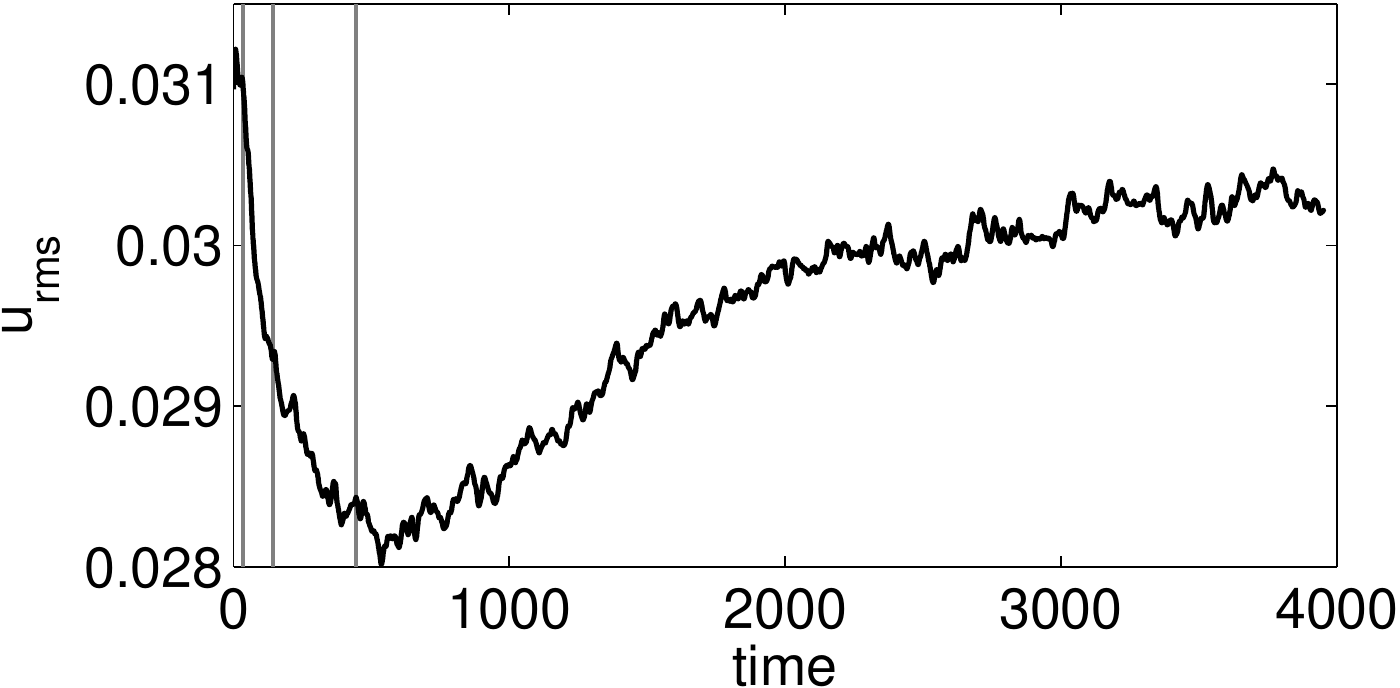}}
  \subfigure[]{
  \includegraphics[clip=true,height=4cm]{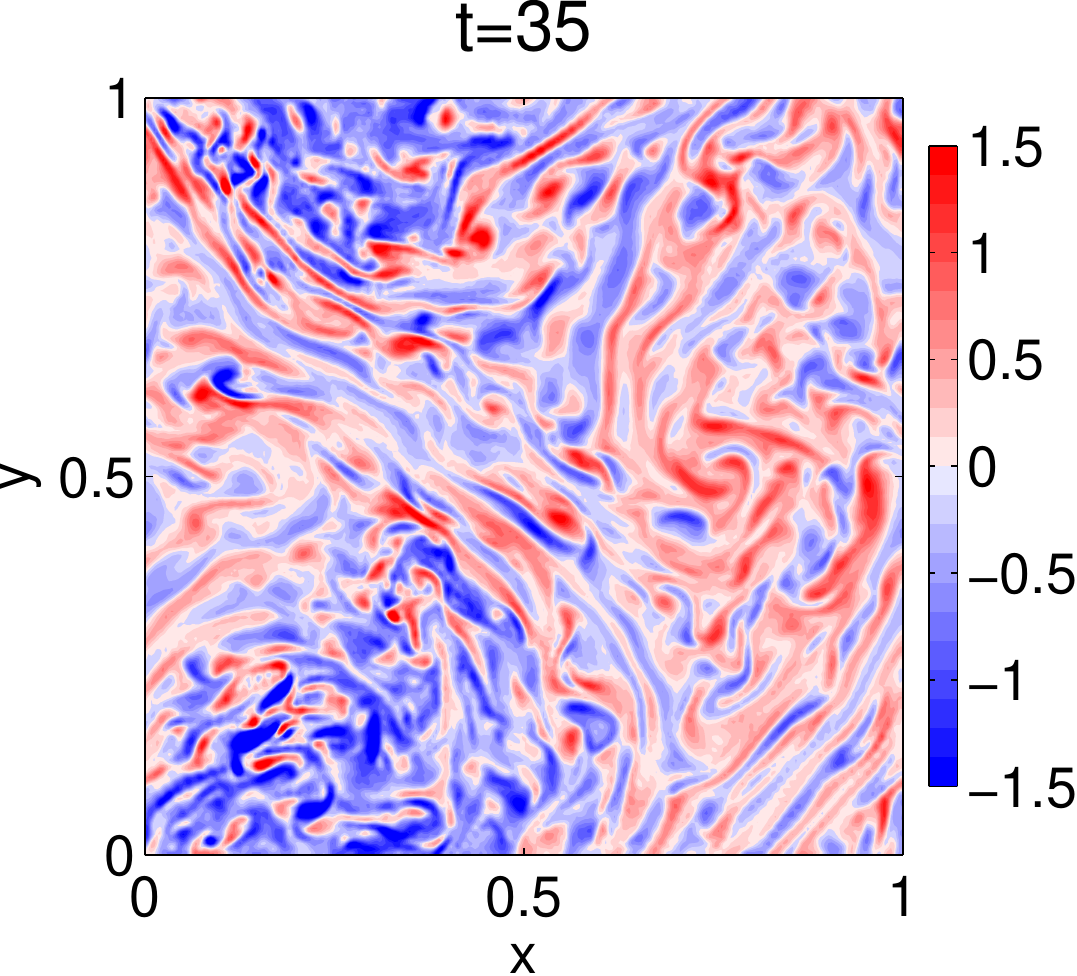}
  \includegraphics[clip=true,height=4cm]{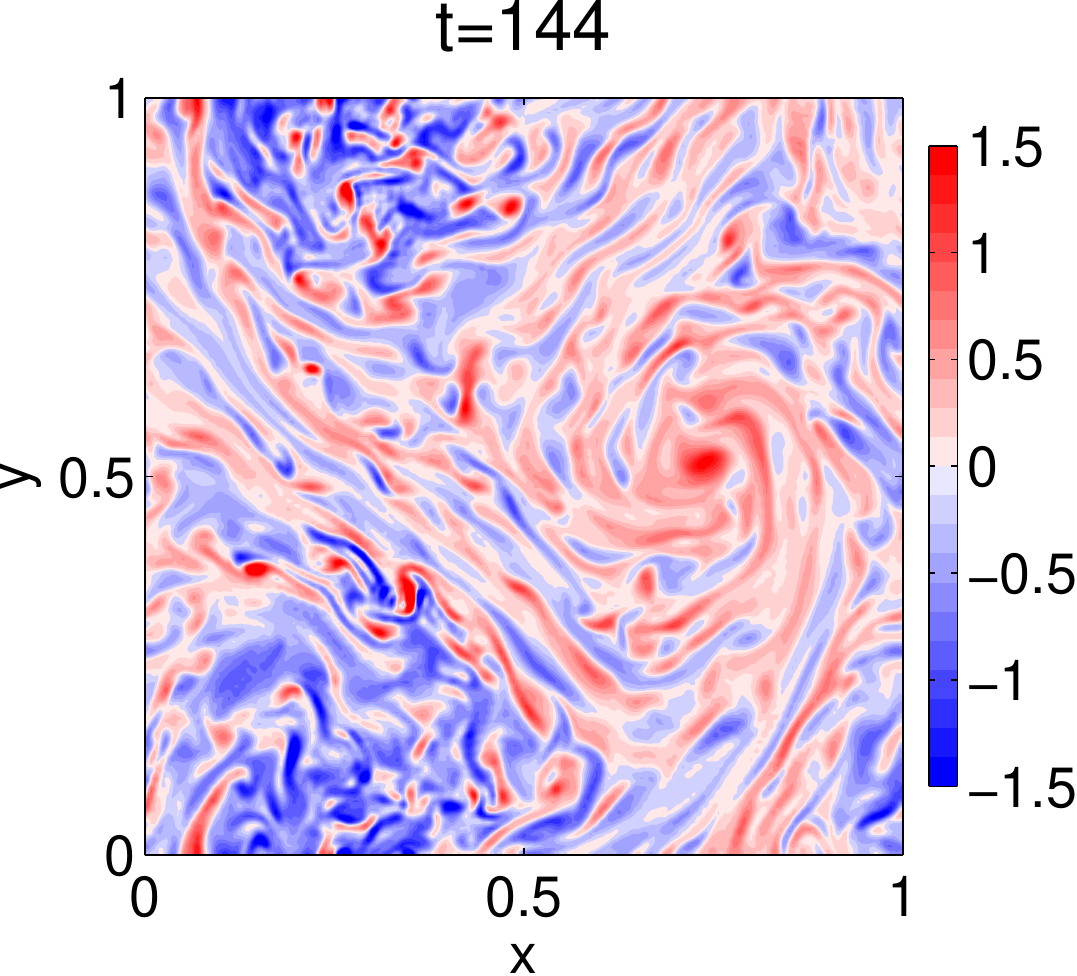}
  \includegraphics[clip=true,height=4cm]{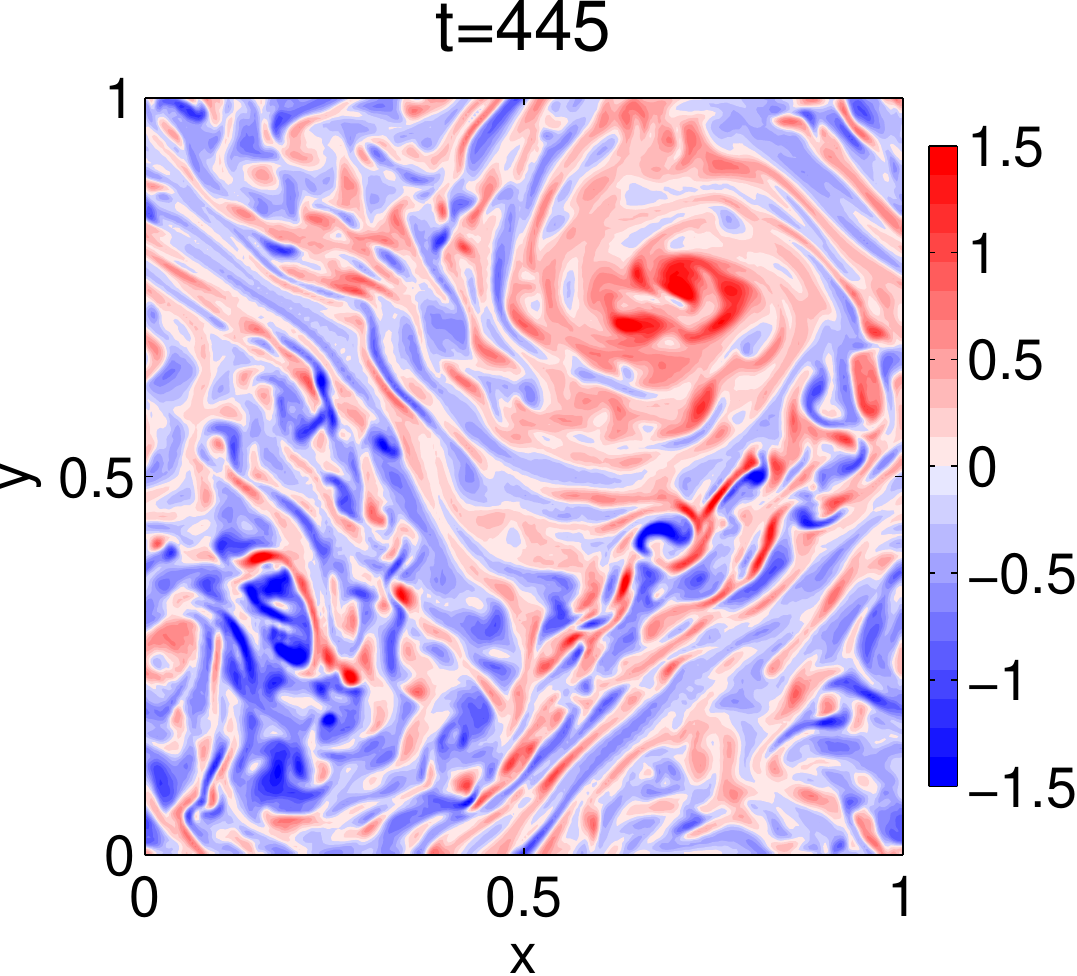}}
  \caption{($a$) Time evolution of the rms velocity after an inversion of the sign
  of the vorticity at $t=0$. 
  ($b$) Horizontal cross-sections (at $z=0.25$) of the axial vorticity
  at different times, indicated by the grey lines in ($a$).
  The simulation was initialised at the time shown in figure~\ref{fig:wz_slice_hcc}, with the same input parameters, 
  but with vorticity of the opposite sign (series~S5, $\tRa=68$).}
\label{fig:wz_invert}
\end{figure}

In order to rule out any dependence of the cyclone/anticyclone asymmetry on initial conditions, 
we restarted a simulation from the snapshot shown in figure~\ref{fig:wz_slice_hcc} with the opposite sign 
for the vorticity (same input parameters otherwise: series~S5, $\tRa=68$). To do so, we changed the sign of the 
three velocity components and the temperature field after first subtracting the horizontal averages. For consistency, 
the horizontal averages of the velocity and temperature fields were kept the same (note that the horizontal average of 
the vertical velocity is zero). Figure~\ref{fig:wz_invert} shows the time evolution of the rms velocity together with 
snapshots of the axial vorticity during the time integration. After a few rotation periods, the concentrated patch of 
anticyclonic vorticity disintegrates into vortices of smaller size. In the meantime, a large-scale cyclonic structure 
emerges through the clustering of smaller size cyclones, increasing in amplitude until a state close to the progenitor 
simulation is reached. The rms velocity decreases rapidly after the start of the simulation at $t=0$, and reaches a 
minimum at $t \approx 500$. At this time, the large-scale cyclone already dominates the horizontal flow. The asymmetry 
between large-scale cyclones and anticyclones is well established after $t \approx 100$, which is about $5-10$ convective 
turnover timescales ($l_h/\langle u_z^2\rangle^{1/2}$). The conclusion to be drawn from this numerical experiment is that 
the cyclone/anticyclone asymmetry is independent of the initial conditions. 

Studies of laboratory and numerical experiments on decaying or forced rotating 3D turbulence report the 
emergence of columnar structures that are predominantly cyclonic \citep[e.g.][]{Hopfinger1982, Bar94, Morize05, Staplehurst2008}. 
In this context, the asymmetry between 
cyclonic and anticyclonic vortices is not yet fully understood \citep{Staplehurst2008}. An argument often quoted 
in the literature is based on the instability of 2D anticyclonic regions that have values of the relative axial 
vorticity close to $-2\Omega$, \ie when the absolute vorticity, $\omega_z + 2\Omega$, is close to zero. In this case, 
the Proudman-Taylor constraint is relaxed, and it is argued that 3D motions can destabilise the 2D anticyclonic structure. 
\citet{Lesieur91} showed that for anticyclones having $|\omega_z| \lesssim \Omega$, the background rotation again 
becomes stabilizing, once the absolute vorticity is significantly larger than the relative vorticity (in absolute value); 
the 2D cyclonic regions, for which the absolute vorticity is always larger  than the relative vorticity, are stabilised by 
the background rotation. In our simulations, the axial vorticity associated with the LSV can reach values locally 
of $2\Omega$ (or greater) for large $\tRa$, so this argument could explain the cyclone/anticyclone asymmetry in these cases. 
However, for simulations with moderate $\tRa$ (for instance for $25 \leq \tRa \leq 51$ in series~S5) and for 
which a large-scale cyclone is formed, the axial vorticity is usually smaller than $2\Omega$; thus this argument does 
not explain the axial vorticity asymmetry in all our cases.

Finally, one further related argument can be invoked to explain the predominance of the large-scale cyclone. 
We noted in \S\,\ref{sec:domain} that the amplitude of the large-scale vortex decays if the local Rossby number, 
$\Ro_z^l$, is greater than about $0.15$. Clearly this implies that a strong effect of rotation on convection 
is necessary for the formation of the large-scale vortex. In an anticyclonic region, the effective rotation 
is weaker than in a cyclonic region; the convection is thus locally less influenced by rotation,
which, in turn, could diminish the degree of anisotropy of the convective structures.

The different arguments presented here to explain the cyclone/anticyclone asymmetry at large scales 
may all act in conjunction or some may prevail for different thermal forcings. The preference for cyclonic vorticity 
induced by the production of intense cyclonic thermal plumes from the thermal boundary layers seems the most convincing 
argument at moderate Rayleigh numbers, where LSV have a small vorticity compared with the planetary vorticity.

Note that in the reduced Boussinesq model of \citet{Jul12} (valid in the small Rossby number limit), the large-scale 
depth-invariant mode consists of a cyclone/anticyclone pair of similar vorticity. In their model, the local vorticity 
is neglected compared with the planetary vorticity in the leading-order equations, so if the vorticity distribution 
is not skewed initially, then the system has no preference for cyclonic or anticyclonic flow. 

The absence of large-scale concentrated anticyclones of the kind observed by \citet{Kap11} and \citet{Chan13} 
is most likely due to the absence of compressibility or stratification in our Boussinesq convection simulations.

\subsection{Transfer of energy to large scales}
\label{sec:filter}

In this subsection, we discuss how energy is transferred from the small scales, where it is injected 
(\ie the horizontal convective size), to the large scales (\ie the box size). In particular, the large-scale flow 
may be the result of an inverse cascade of energy, similar to 2D turbulence \citep[e.g.][]{Bof12}, where the energy 
is transferred to the smallest wavenumber across the whole spectrum, or of a direct transfer from the combination of 
two small-scale modes of comparable wavenumber as, for instance, in a mean-field instability \citep[e.g.][]{Frisch1987}.

The equation for the evolution of the axial vorticity is
\begin{eqnarray}
  		\frac{\partial \omega_z}{\partial t} + ( \vel \bcdot \bnabla )  \omega_z  
            	= (2\Omega +\omega_z)  \frac{\partial u_z}{\partial z}
               	+ (\vor_h \bcdot \bnabla ) u_z
               	+ \nu \nabla^2 \omega_z ,
\label{eq:omegaz}
\end{eqnarray}
where $\vor_h = (\omega_x,\omega_y,0)$. Equation~(\ref{eq:omegaz}) shows that $z$-invariant vortical flow can be 
produced only by nonlinear interactions, since the $z$-average of the vortex stretching term \mbox{$2\Omega \partial_z u_z$} 
is zero owing to the impenetrable boundary conditions.

\begin{figure}
\centering
  \subfigure[]{
  \raisebox{-0.1cm}{\includegraphics[clip=true,height=5.8cm]{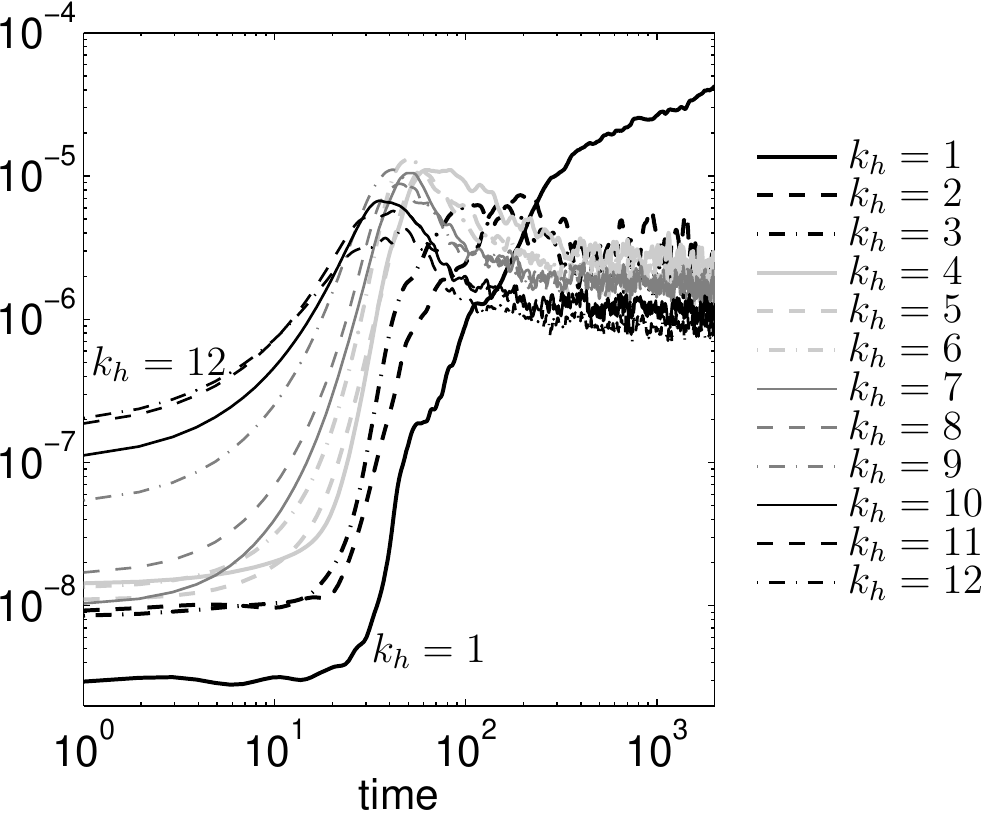}}}
  \subfigure[]{
  \includegraphics[clip=true,height=5.5cm]{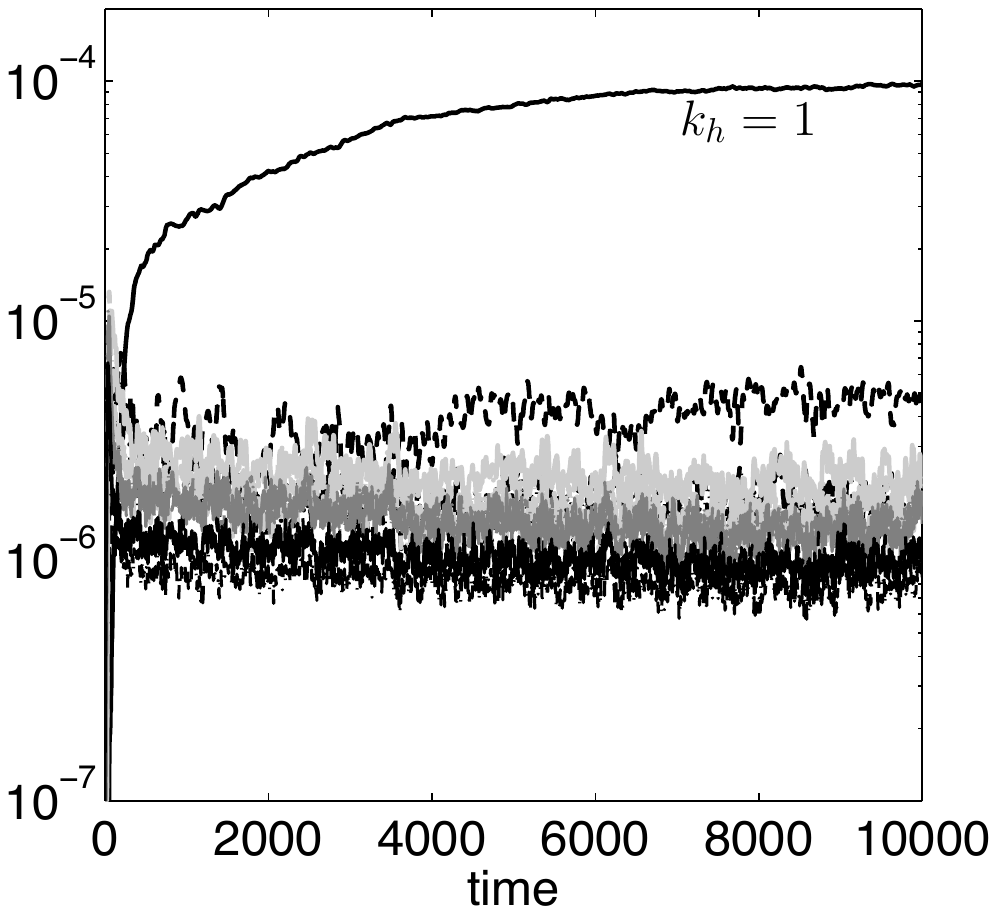}}
  \caption{Time series of the kinetic energy in each horizontal wavenumber $1\leq k_h \leq 12$
  for $\tRa=34$ of series~S5: ($a$) initial growth using a logarithmic scale for time; ($b$) 
long time integration using a linear scale. The initial condition was a snapshot of a simulation with the same 
parameters except that $\tRa=13$.}
\label{fig:EnergiesModes_Ta4e10Ra4e8}
\end{figure}

To gain some insight into the mechanism of formation of the LSV, we study the behaviour of the kinetic energy 
of different horizontal modes, first as a function of time, and then as a function of increasing Reynolds number. 

Figure~\ref{fig:EnergiesModes_Ta4e10Ra4e8} shows time series of the kinetic energy contained in each horizontal 
wavenumber $k_h \in [1,12]$, summed over all vertical wavenumbers $k_z$, for $\tRa=34$ of series~S5. The starting point 
was a prior simulation performed just above the onset of convection for this series ($\tRa=13$), where $k_h=12$ is the 
dominant wavenumber; at time $t=0$ the Rayleigh number was increased to $\tRa=34$. The modes $k_h\leq 4$ are linearly stable 
to convection for this Rayleigh number, so they grow only once the nonlinear interactions of larger wavenumbers gain 
sufficient amplitude, after $t \approx 20$. For $30 \lesssim t \lesssim 50$, the modes $k_h=1 \textrm{ -- } 3$ have a 
roughly similar growth rate, which is larger than that of modes close to the marginally stable mode at onset 
($k_h=10 \textrm{ -- } 12$). As the modes $k_h>4$ saturate for $t \gtrsim 50$, the large-scale modes $k_h=1 \textrm{ --} 3$ 
carry on growing but with a diminished growth rate. Eventually, for $t \gtrsim 200$, the modes $k_h = 2 \textrm{ -- } 3$ 
saturate at a greater amplitude than that of the larger wavenumbers. However, the mode $k_h=1$ continues to grow at a yet 
smaller growth rate, about \mbox{$0.0004$}. It eventually saturates for times $t \gtrsim 10^4$. The final slowly 
growing phase of the $k_h=1$ mode is the process that we aim to understand in the remainder of this section.

\begin{figure}
\centering
 \subfigure[horizontal flow]{
  \includegraphics[clip=true,width=5cm]{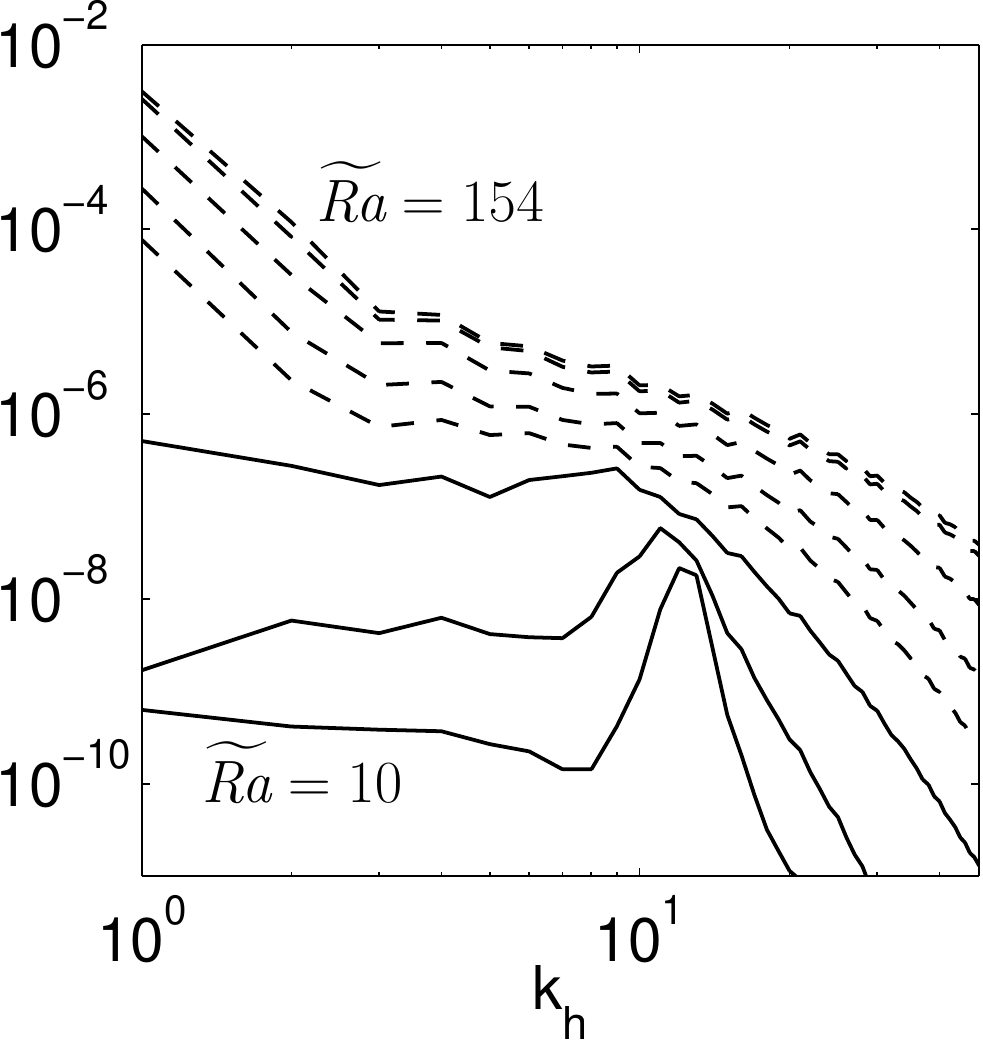}}
   \subfigure[vertical flow]{
  \includegraphics[clip=true,width=5cm]{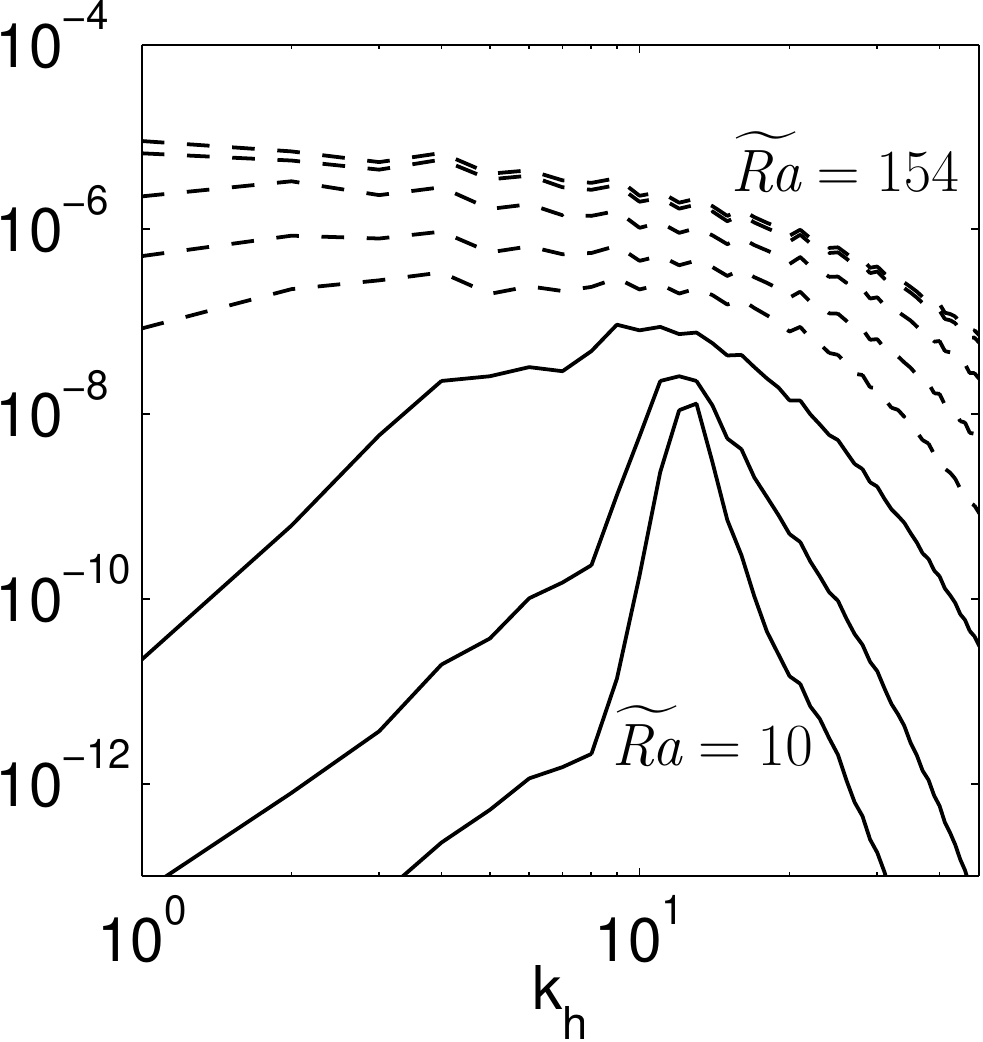}}
  \caption{Kinetic energy spectra in the horizontal directions of the horizontal and vertical
velocities for different Rayleigh numbers for series~S5. The solid lines
correspond to $\tRa = 10$, $13$, $20$, for which $\Gamma\approx 1$,
and the dashed lines $\tRa=34$, $51$, $86$, $137$, $154$, for which $\Gamma >1$.
The spectra are calculated from equations~(\ref{eq:Eh}) and~(\ref{eq:Ev}).}
\label{fig:spectrum_Ta4e10}
\end{figure}

\begin{figure}
\centering
  \subfigure[]{
  \includegraphics[clip=true,height=5cm]{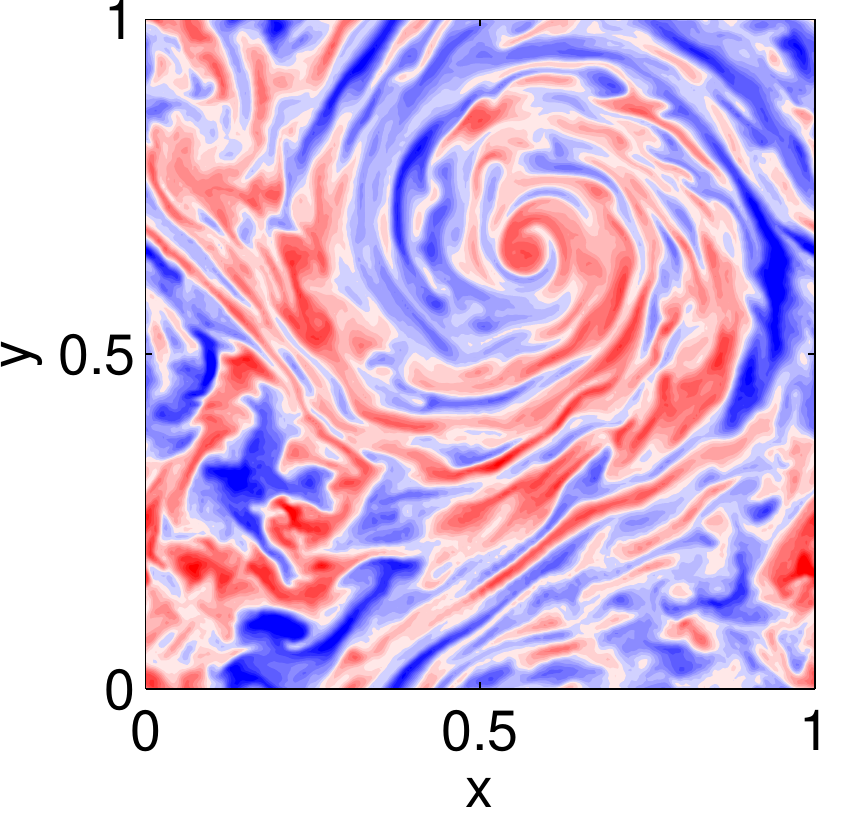}}
  \subfigure[]{
  \includegraphics[clip=true,height=5cm]{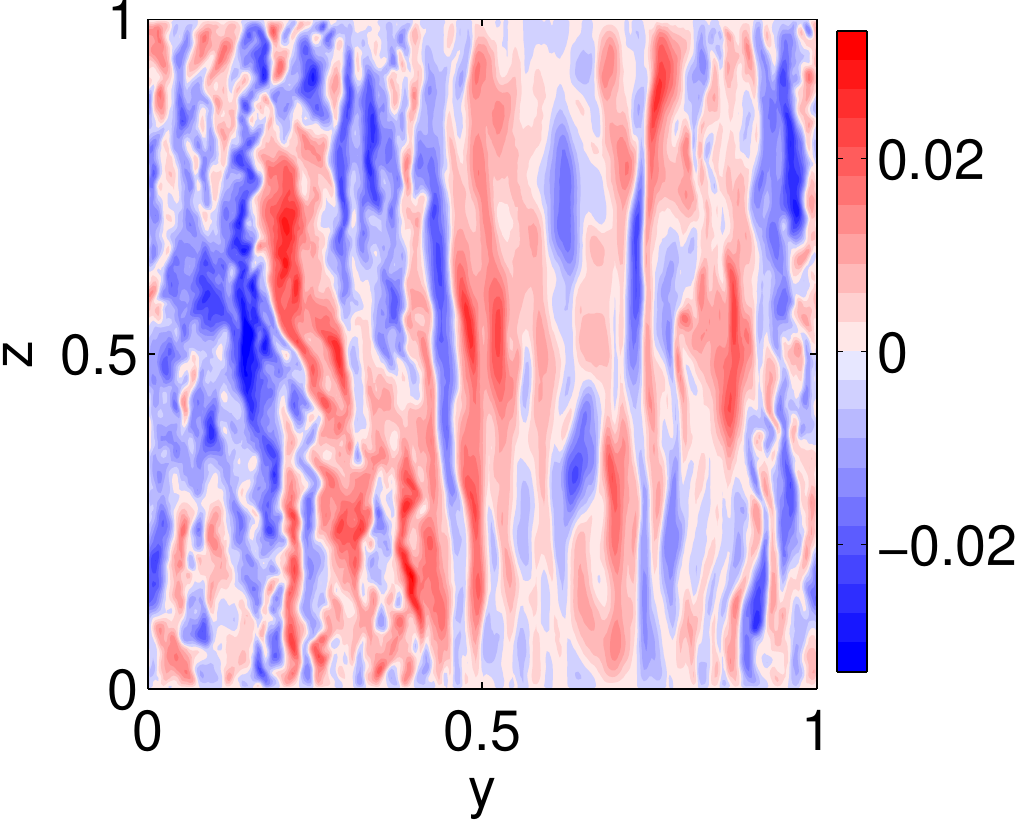}}
  \caption{Cross-sections of the vertical velocity in ($a$) a horizontal plane at $z=0.25$ and 
($b$) a vertical plane at $x=0.6$; $\tRa=154$ of the series~S5.}
\label{fig:uz_Ta4e10Ra1.8e9}
\end{figure}

The behaviour of the kinetic energy spectrum as the Reynolds number increases can be studied in 
figure~\ref{fig:spectrum_Ta4e10}, where we plot the kinetic energy spectra of the horizontal and 
vertical velocities for the series~S5. For the horizontal flow, energy is transferred from the convective 
scale to larger scales even when $\Gamma \approx 1$, but the spectra steepen significantly at large scales 
when $\Gamma>1$. For cases with $\Gamma>1$, the slope is larger for $k_h=1 \textrm{ -- 2}$ than for 
$k_h = 3 \textrm{ -- } 10$, thereby indicating that the kinetic energy accumulates at the smallest available wavenumber. 
The vertical kinetic energy also undergoes a progressive transfer to larger scales for increasing $\tRa$. For $\tRa\leq 86$, 
this transfer occurs mainly to the benefit of the modes $2\leq k_h \leq 12$, and for $\tRa\geq 137$ the vertical 
velocity is eventually dominated by the mode $k_h=1$. For large $\tRa$, the presence of large-scale vertical velocities 
could be due to a secondary recirculation associated with the LSV. In this case, by continuity, the large-scale vertical 
flows would be expected to be antisymmetric with respect to the horizontal mid-plane, since the horizontal flows of the 
LSV are largely $z$-invariant, whereas buoyancy-driven vertical motions are mostly symmetric with respect to the mid-plane. 
Figure~\ref{fig:uz_Ta4e10Ra1.8e9} shows cross-sections of $u_z$ for the case $\tRa=154$ of figure~\ref{fig:spectrum_Ta4e10}. 
In the horizontal cross-section, the isocontours of $u_z$ are elongated horizontally. 
In the vertical cross-section, which is taken through the large-scale cyclone, the vertical velocity is mostly symmetric 
with respect to the mid-plane. This indicates that vertical motions are essentially driven by the buoyancy force rather 
than by a recirculation of fluid associated with LSV. The transfer to large scale in the vertical kinetic energy is 
therefore most likely due to the reorganization of the convection by the LSV, which leads to horizontally elongated structures. 

\begin{table}
\centering
\small
\begin{tabular}{c  c  c  c  c  c  c  c  c  c}
\hline
Case: & A & B & C & D & E & F & G & H & I \vspace{0.3cm}
\\ 
$k_z=0$ & $1$, $11-13$ & $1$, $6-8$ & $1$, $4-6$, $11-13$ & $1$ & $1$ & $1$ & $1$ & $1$ & $1$ 
\vspace{0.3cm}
\\
$k_z \ne 0$ & $1$, $11-13$ & $1$, $6-8$ & $1$, $4-6$, $11-13$ & all & $\geq 2 $ & $\geq 6$ & $\geq 12$ & $\geq 15$ & $\geq 21$
\vspace{0.3cm}
\\
$\frac{E(k_h=1)}{E (k_h\geq 1)}$ & $0.02$ & $0.05$ & $0.03$ & $0.77$ & $0.69$ & $0.91$ & $0.86$ & $0.73$ & $0.03$
\\ \hline
\end{tabular}
\caption{Summary of the filtered simulations. The entries of lines $2$--$3$ in the table refer 
to the horizontal wavenumbers that are retained during the simulation. 
The entries of line $4$ denote the ratio of the kinetic energy in $k_h=1$ to the total kinetic energy 
calculated at the end of the simulation. For comparison, in the full simulation the ratio is about $0.81$.}
\label{tab:Filter}
\end{table}

\begin{figure}
\centering
  \subfigure[Case A]{\label{fig:filter_k12}
  \includegraphics[clip=true,height=5.7cm]{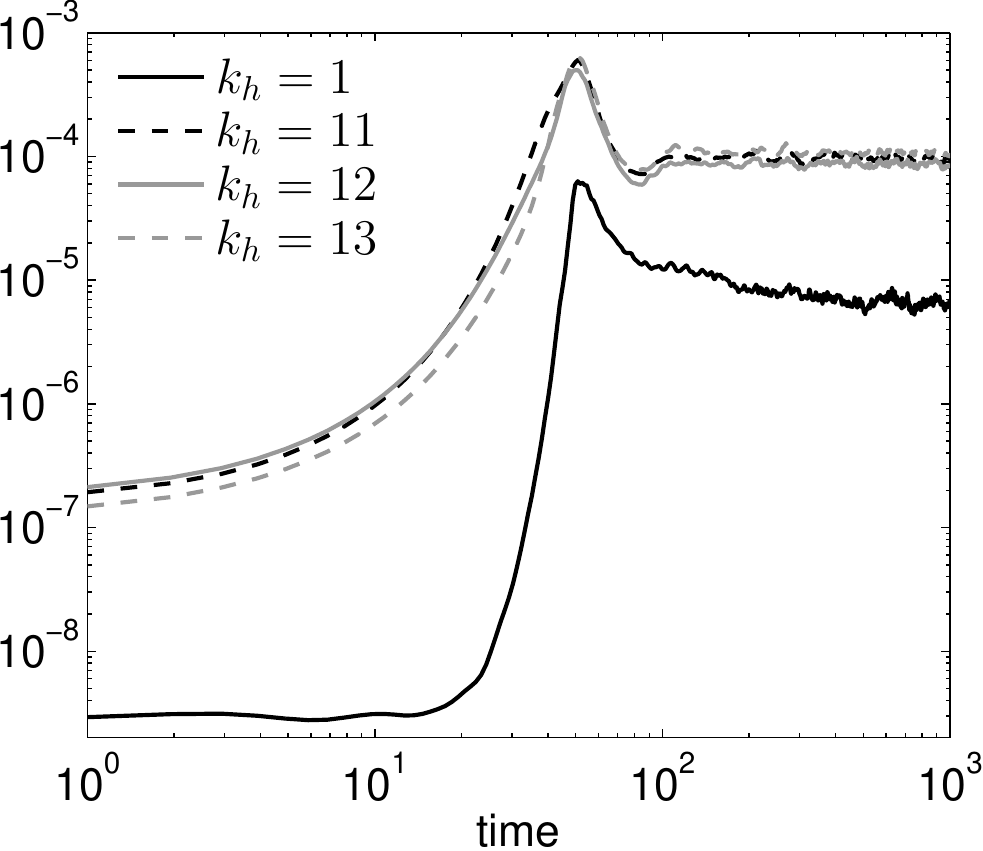}}
  \subfigure[kinetic energy in $k_h=1$]{\label{fig:filter_kh1}
  \includegraphics[clip=true,height=5.5cm]{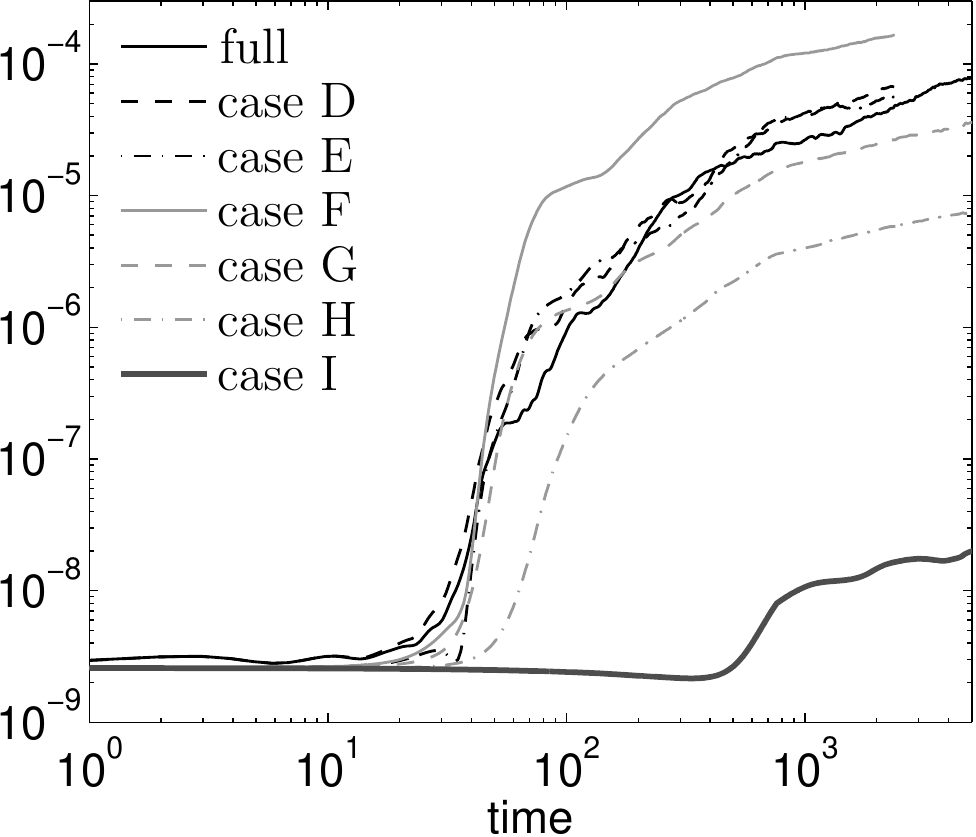}}
  \caption{($a$) Time series of the kinetic energy for each $k_h$ in case~A.
	   ($b$) Time series of the kinetic energy of the $k_h=1$ mode for several different filtered 
simulations, summarized in table~\ref{tab:Filter}.}
\end{figure}

To determine the ranges of wavenumbers that contribute to the transfer of energy to the large 
horizontal scale, we performed a series of numerical simulations in which a given range of horizontal and 
vertical wavenumbers of the flow is filtered, \ie set to zero throughout the time integration. In all 
the filtered simulations, the input parameters are set to $\tRa=34$ of series~S5 and the initial condition 
is the same as in figure~\ref{fig:EnergiesModes_Ta4e10Ra4e8}. Table~\ref{tab:Filter} summarises the different cases. 
First, to test whether a narrow range of spectral modes can sustain a significant $k_h=1$ mode, we considered cases 
where only the modes $k_h\in [11 \textrm{ -- } 13]$ (case~A),  $k_h \in [6  \textrm{ -- } 8]$ (B) and 
$k_h \in [4  \textrm{ -- } 6, 11  \textrm{ -- } 13]$ (C) are retained in addition to $k_h=1$. 
(Recall that $k_h = \left({k_x^2+k_y^2}\right)^{1/2}$; the $k_h = n$ bin includes all modes 
in the range $n-1/2 \leq k_h < n+1/2$.)
For these modes, all of the vertical wavenumbers are retained. Figure~\ref{fig:filter_k12} shows the 
kinetic energy for each $k_h$ in case~A. The same two initial phases for the evolution of $k_h=1$ can be observed 
as in the full simulation (figure~\ref{fig:EnergiesModes_Ta4e10Ra4e8}): no growth at first ($0 \lesssim t \lesssim 20$), 
followed by a rapid growth ($30 \lesssim t \lesssim 50$) when the small-scale modes are large enough to provide a 
significant amplitude to the nonlinear terms fuelling $k_h=1$. However, the subsequent evolution of $k_h=1$ is different 
from that of the full simulation; as the larger wavenumbers saturate, the kinetic energy of $k_h=1$ also saturates at a 
much smaller value. Cases~B and~C exhibit a very similar evolution for $k_h=1$. These three numerical experiments demonstrate 
that a narrow range of wavenumbers in spectral space cannot directly produce a large-scale flow of significant amplitude, 
even when an intermediate range of horizontal wavenumbers is retained (case~C).

To determine if the energy transfer from small to large scales is due mostly to the interaction of strictly 
$z$-invariant modes (as in 2D turbulence) or to the interaction of $z$-dependent modes, we performed a series 
of filtered simulations in which the vertical wavenumber $k_z=0$ is suppressed for all $k_h$ except $k_h=1$ 
(cases~D--I). Figure~\ref{fig:filter_kh1} shows the kinetic energy in mode $k_h=1$ for the different cases. 
We consider 
whether the $k_h=1$ mode reaches an amplitude significantly higher than 
those of the other horizontal wavenumbers and has a temporal evolution similar to that of the full simulation. 
Table~\ref{tab:Filter} contains the ratio of the kinetic energy in $k_h=1$ to the total kinetic energy at the 
end of the simulations. For case~D, where all the $k_z\ne 0$ horizontal modes are retained, a \mbox{$k_h=1$} 
mode of large amplitude is produced. For cases~E--I, we further filter the flow by suppressing all the intermediate 
modes \mbox{$1<k_h<k_h^f$}, where $k_h^f$ is a given mode. For $k_h^{f} \leq 15$ (cases~E--H), a mode \mbox{$(k_h,k_z)=(1,0)$} 
of large amplitude is still generated. However for $k_h^{f}=21$ (case~I), this large-scale mode has only a very weak amplitude. 
These numerical experiments indicate that the generation of the large-scale $z$-invariant mode does not necessarily require 
the interaction of $z$-invariant modes, \ie it is not the product of a purely 2D inverse cascade. They further suggest 
that the generation of the large-scale mode does not require the intermediate wavenumbers; thus it is not produced by 
an inverse energy cascade from interactions of $z$-dependent modes, but by a direct transfer from interactions 
of $z$-dependent modes at horizontal scales close to the convective scale.

Using the reduced model of Boussinesq convection of \citet{Jul12}, \citet{Rubio13} calculate the transfer 
functions of the kinetic energy of the $z$-invariant modes when the LSV are present. They find that the forcing 
produced by nonlinear interactions of $z$-dependent modes with $k_h \geq 8$ transfers energy directly to the 
large-scale $z$-invariant mode. This observation is in agreement with the result of our filtered simulations. 
However, \citeauthor{Rubio13} find that this process occurs only during the slow growth of the mode $k_h=1$, when 
its energy is dominant. During earlier stages of the time evolution, they argue that the action of $z$-invariant 
flows on the $z$-dependent eddies leads to increased coherence of the $z$-dependent forcing. This result contradicts 
somewhat the results from our filtered simulations, where we find that in the 
absence of $z$-invariant modes other than $k_h=1$, a large-scale $z$-invariant mode still grows to become dominant. 
However, it is possible that the increased coherence of the $z$-dependent forcing is due only to the action of 
the $z$-invariant $k_h=1$ mode.

\subsection{Effect on the heat transfer}
\label{sec:heat}
In simulations of compressible convection in Cartesian domains, \citet{Kap11} and \citet{Chan13} observe that 
the LSV are associated with temperature anomalies; the central parts of the cyclones (anticyclones) are colder 
(warmer) than their surroundings over most of the vertical extent of the convective layer. The increase of the effective 
rotation in cyclones means that convection is further inhibited by rotation, and it is suggested that this leads to further 
cooling of the region. By contrast, in anticyclones, the Proudman-Taylor constraint is relaxed, and so convection can 
develop more efficiently, which is interpreted as the warming of the region.  Note that in these compressible models, 
the convective layer is not necessarily in direct contact with the top and bottom boundaries; a convectively stable 
layer is added either above the convective layer \citep{Chan13} or above and below the layer \citep{Kap11}. A further 
difference to our set-up is that these models also employ mixed temperature boundary conditions, with fixed flux at 
the bottom and fixed temperature at the top.
 
\begin{figure}
\centering
  \subfigure[]{\label{fig:heatflux_snap}
  \includegraphics[clip=true,width=5.5cm]{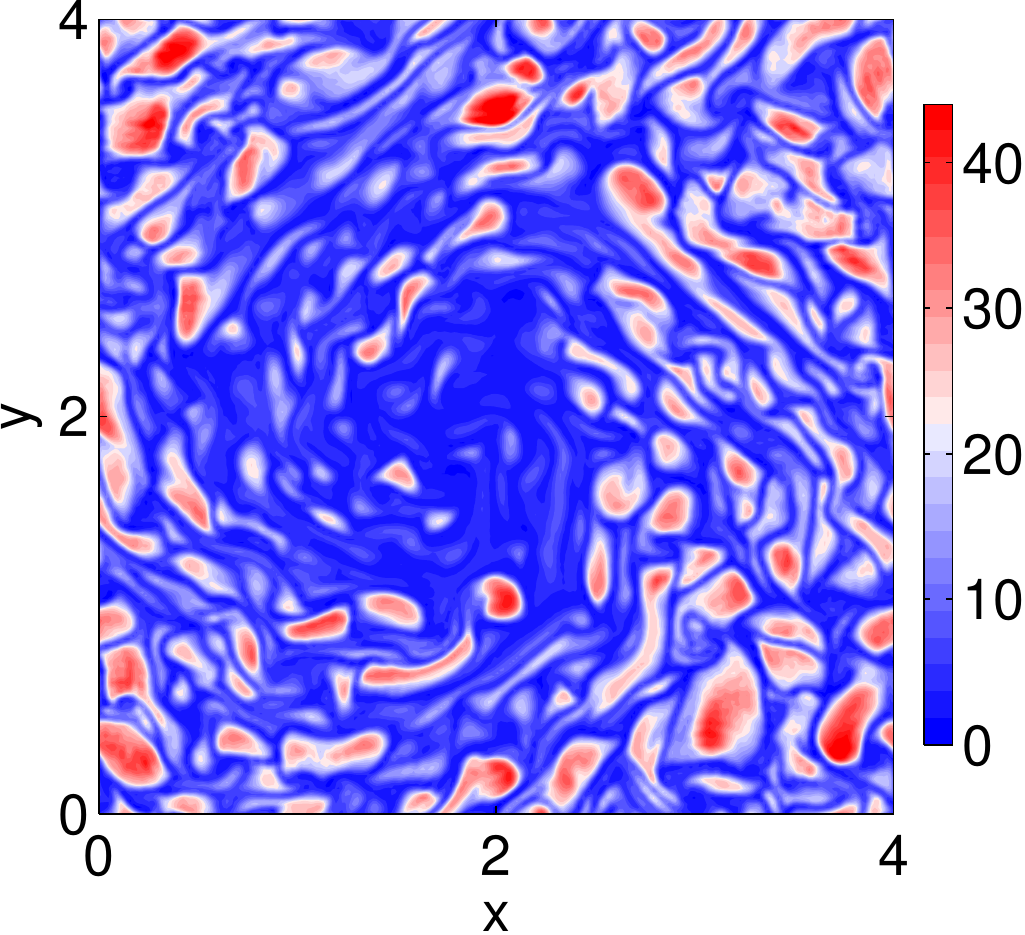}}
  \subfigure[]{\label{fig:heatflux_ave}
  \includegraphics[clip=true,width=5.5cm]{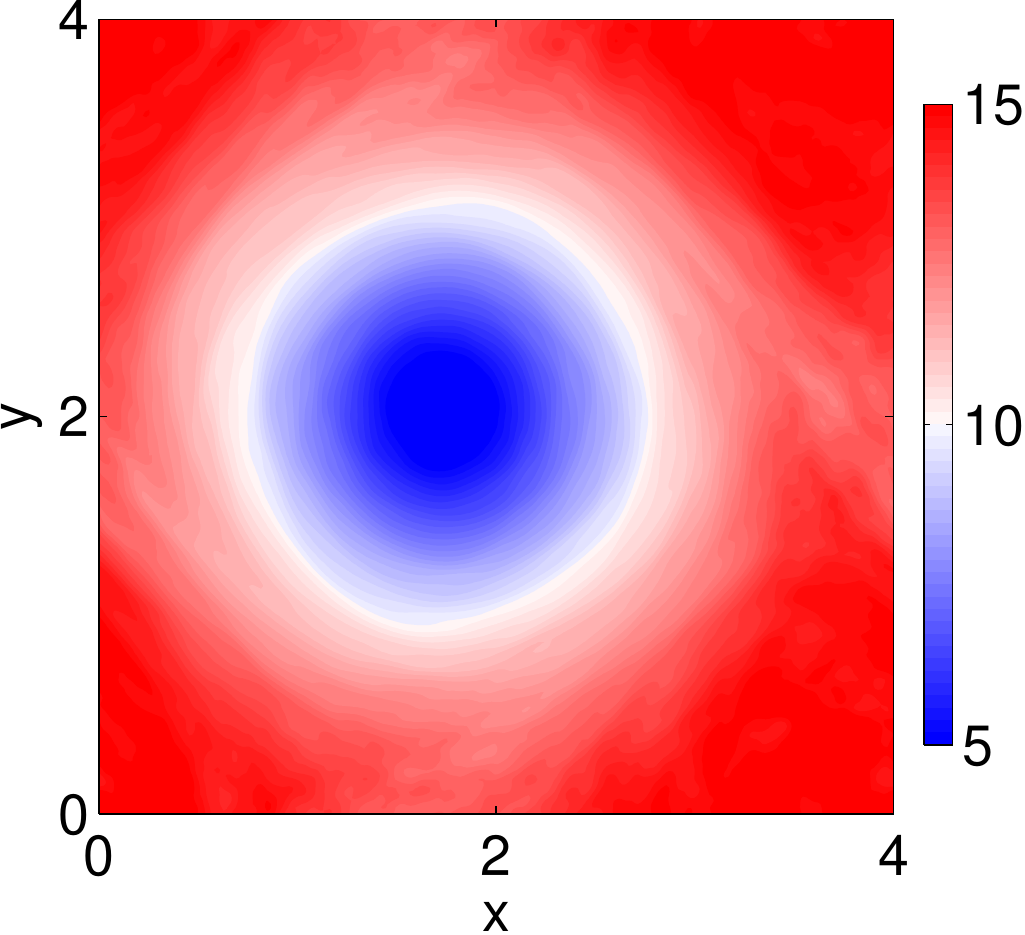}}
  \subfigure[]{\label{fig:wz_ave}
  \includegraphics[clip=true,width=5.8cm]{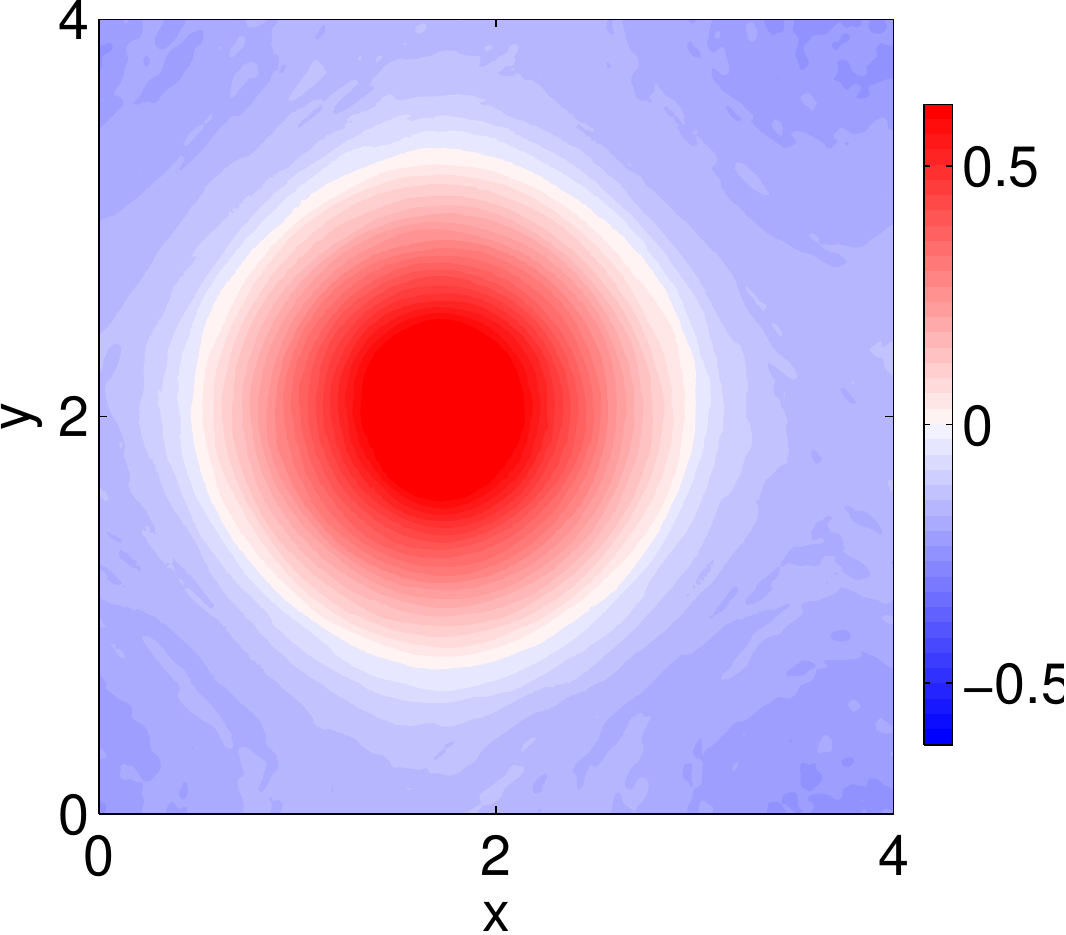}}
  \subfigure[]{\label{fig:Tave_profile}
  \includegraphics[clip=true,width=5cm]{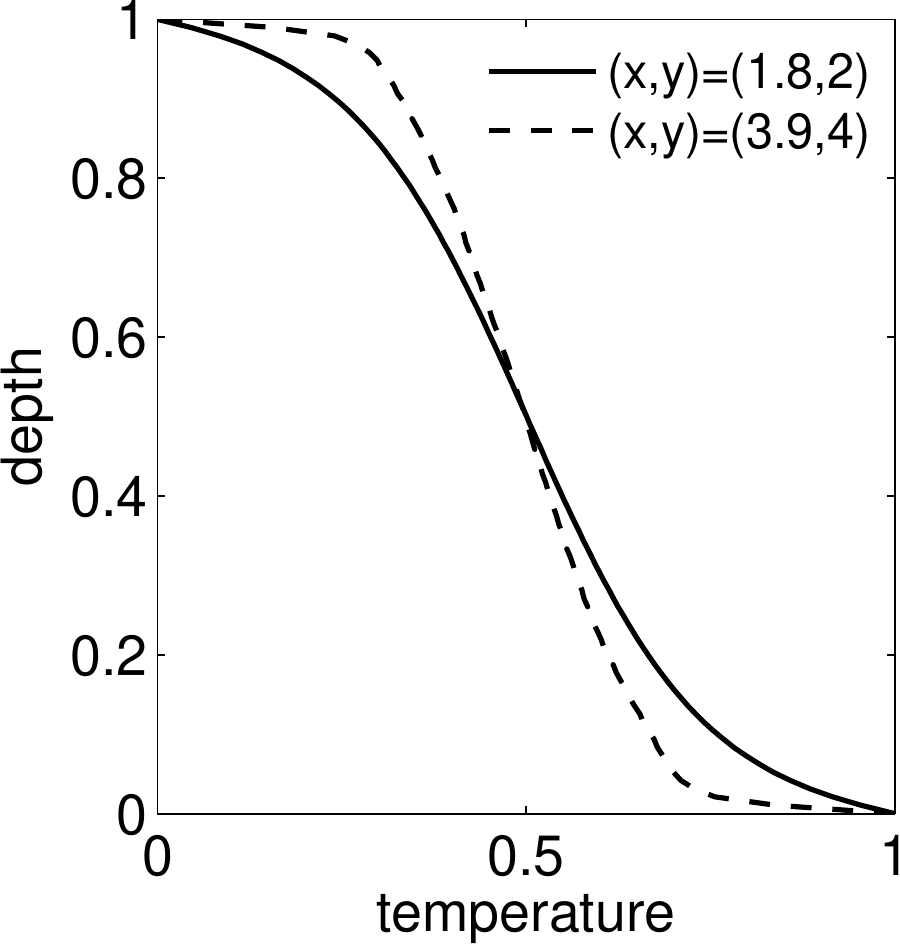}}
  \caption{($a$) Snapshot of the heat flux at the upper boundary, $z=1$.
  ($b$) Time-average of the heat flux at $z=1$. ($c$) Time- and vertically-averaged axial vorticity.
  The cyclone stays roughly at the same location during the time average. 
($d$) Vertical profile of the time-averaged temperature added to the linear background profile, $1-z$, 
inside the core of the cyclone ($(x,y)=(1.8,2)$) and in the anticyclonic region ($(x,y)=(3.9,4)$). 
Parameters: $\tRa=46$ of series~S3.}
\end{figure}

In a Boussinesq system, vertical temperature profiles are expected to be symmetric with respect to the horizontal 
mid-plane on average. Instead of the temperature, we therefore examine the heat flux at the upper surface ($z=1$), defined as
\begin{equation}
 q = - \left. \frac{\partial \theta}{\partial z} \right|_{z=1} +1 .
\end{equation}
Note that the contribution from the linear temperature background, $1-z$, is included in the definition of $q$. 
Figure~\ref{fig:heatflux_snap} shows the instantaneous heat flux at the upper surface for the case $\tRa=46$ of 
the series~S3, where a cyclone of large amplitude is present. In the central region of the horizontal plane, which 
is located just above the core of the cyclone, no patches of large heat flux are present, unlike in the surroundings. 
During this simulation, long time-averages of $q$ (shown in figure~\ref{fig:heatflux_ave}) and of the $z$-averaged 
axial vorticity (figure~\ref{fig:wz_ave}) are calculated. The time-average is taken during a period when the cyclone 
remains roughly in the same location in a horizontal plane. The averaging process reveals a distinct patch of weak 
heat flux above the cyclone, with the minimum in the heat flux about three times smaller than the maximum.

Figure~\ref{fig:Tave_profile} shows vertical profiles of the time-averaged temperature (including 
the linear background profile) both inside the core of the cyclone, at $(x,y)=(1.8,2)$, and in its surroundings, 
in the weaker large-scale anticyclonic circulation, at $(x,y)=(3.9,4)$. The vertical temperature profile is less 
steep in the core of the cyclone in the bulk of the fluid compared with the profile in the anticyclonic region. 
This implies that the vertical mixing of temperature is less efficient inside the cyclone, possibly as a consequence 
of the local increase of the rotation, thus inhibiting convection. The thermal boundary layers, where the vertical 
temperature gradient is larger than in the bulk, are thinner in the anticyclonic region than in the core of the cyclone, 
which explains the heat flux anomaly at the upper surface. 

\begin{figure}
\centering
  \includegraphics[clip=true,width=7cm]{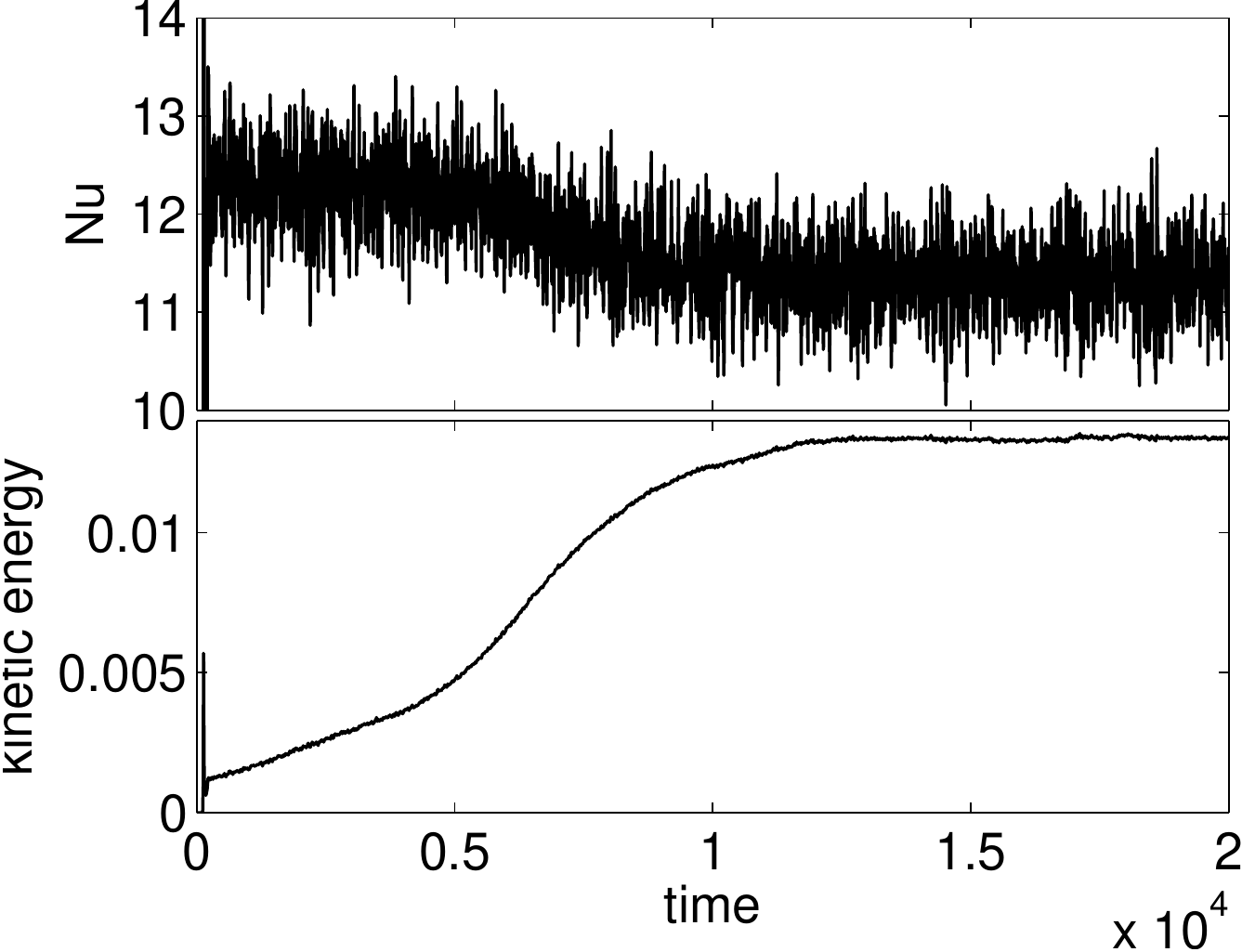}
  \caption{Time series of the Nusselt number and the kinetic energy for $\tRa=37$ of series~S3.}
\label{fig:Effect_Nu}
\end{figure}

Since the large-scale cyclonic structure disturbs both the convective structures and the heat flux, 
we might expect a reduction of the efficiency of vertical convective transport. The efficiency of the heat 
transfer is usually quantified by the Nusselt number, $\Nu$, which is a measure of the total heat flux through 
the layer normalised by the heat flux in the absence of convective motions. Figure~\ref{fig:Effect_Nu} shows 
time series of the Nusselt number and the kinetic energy for the simulation shown in 
figure~\ref{fig:wz_Ta1e8Ra8e6_l4} ($\tRa=37$ in the series~S3). When the convection is initially established 
for $t \lesssim 5000$, the mean Nusselt number is about $12.3$. $\Nu$ then decreases when the large-scale circulation 
grows significantly for $t \gtrsim 5000$. As the kinetic energy saturates, $\Nu$ eventually 
reaches a mean value of $11.3$, roughly $8$\% smaller than the initial $\Nu$.

\begin{figure}
\centering
  \includegraphics[clip=true,width=7cm]{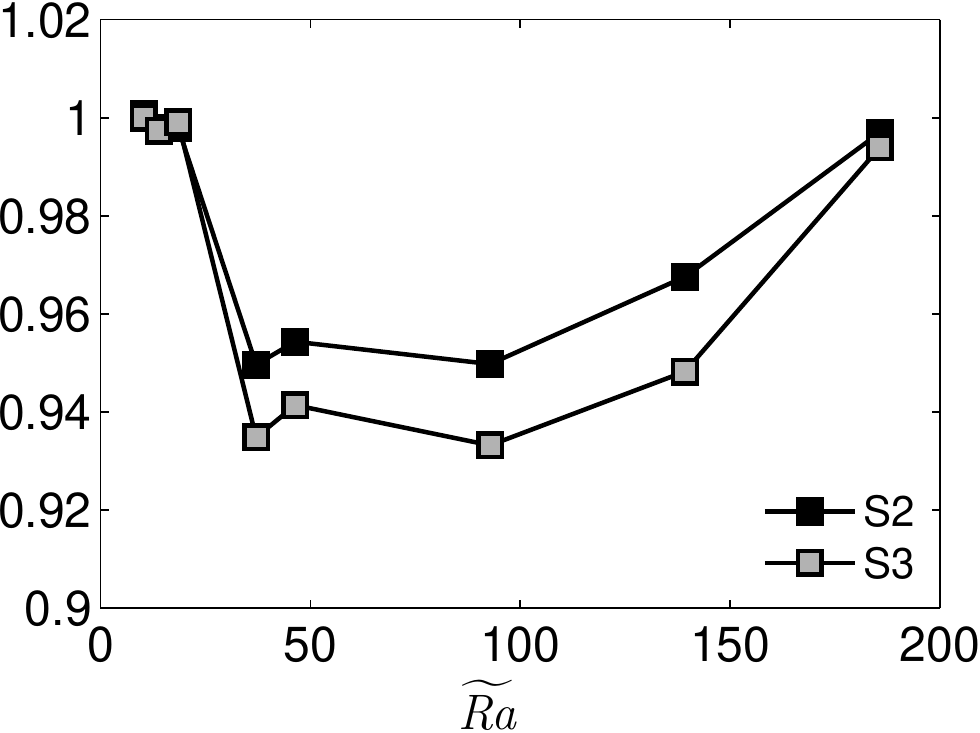}
  \caption{$\RNu$ as a function of $\tRa$. $\RNu$ is the ratio of the 
  Nusselt number in series~S2 and S3
  to the Nusselt number in the series~S1 for the same $\tRa$.}
\label{fig:ratio_Nu}
\end{figure}

A systematic evaluation of the decrease of the Nusselt number in the presence of LSV compared 
with its value when the convection is first becoming established (for example at times $t \lesssim 5000$ 
in figure~\ref{fig:Effect_Nu}) is not possible for most of our simulations, since we generally use a snapshot 
of a simulation at smaller Rayleigh number as the starting point for a new simulation. Instead, we calculate $\RNu$, 
the ratio of the Nusselt number measured in the saturated phase in the series~S2 ($\aspect=2$) or S3 ($\aspect=4$) 
to the Nusselt number in the series~S1 ($\aspect=1$) for the same $\Ra$, the three series having the same Ekman number. 
The large-scale horizontal flows maintain an amplitude close to the vertical flows in series~S1 ($\Gamma$ remains close to 1), 
so it is reasonable to assume that the Nusselt number is unaffected by the presence of the large-scale horizontal flows 
in this series. $\RNu$ is plotted in figure~\ref{fig:ratio_Nu} as a function of $\tRa$. In the series~S2 and S3, $\RNu$ 
is about unity when $\tRa \lesssim  20$, as expected in the absence of LSV. For $20 \lesssim \tRa \lesssim 100$, $\RNu$ is 
about $0.95$ for the series~S2 and $0.93$--$0.94$ for the series~S3. The value of $\RNu$ for $\tRa=37$ of series~S3 is 
consistent with the observation made in figure~\ref{fig:Effect_Nu}. Although $\Gamma$ increases with $\tRa$ within this 
range of $\tRa$, the decrease of the Nusselt number is always about $5\%$ for series~S2 and $6$--$7\%$ for series~S3. 
For $\tRa \gtrsim 100$, $\RNu$ tends to $1$, which is consistent with the decrease of $\Gamma$ in series~S2--S3.

With their reduced model, \citet{Jul12} observe that the growth of a large-scale cyclone and anticyclone pair is, 
on the contrary, accompanied by an \textit{increase} in the Nusselt number. Since their model does not possess the 
cyclone/anticyclone asymmetry, it is perhaps not surprising that the effect of the LSV on the heat transfer is different 
in the reduced model and in our 3D model. However, the explanation for the increase of the Nusselt number in the study 
of \citeauthor{Jul12} remains unclear.

\begin{figure}
\centering
  \includegraphics[clip=true,width=8cm]{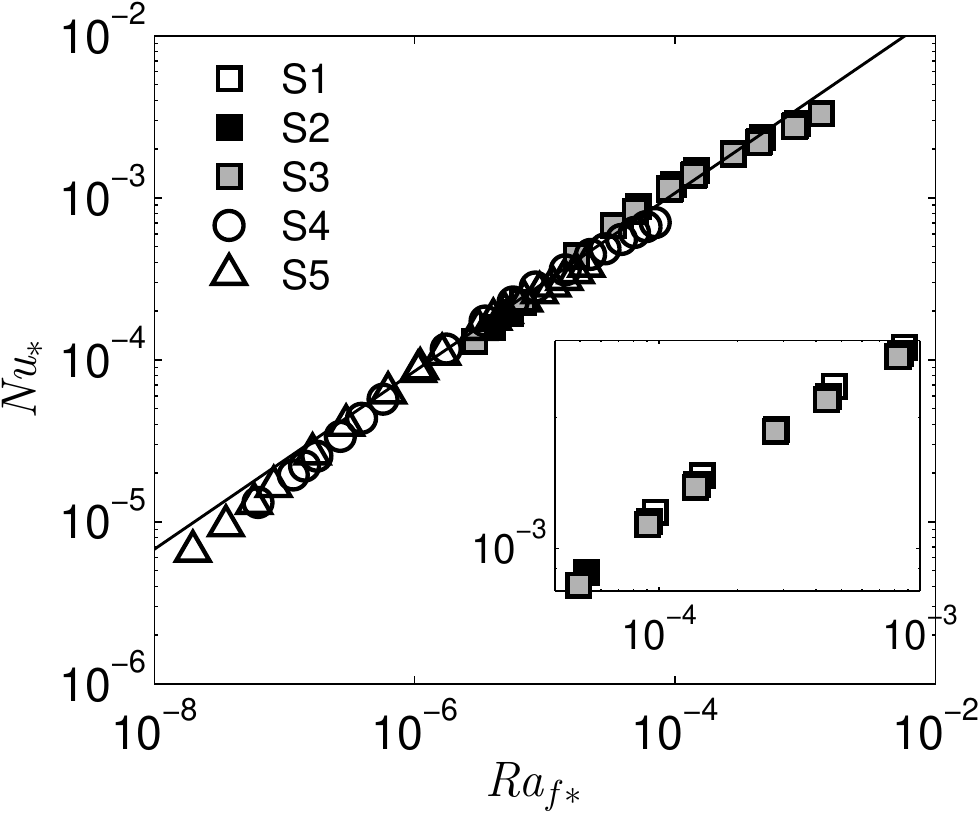}
  \caption{$\Nu_{\ast}$ as a function of $\Ra_{f\ast}$. 
  The solid line corresponds to $\Nu_{\ast} = 0.17 \Ra_{f\ast}^{0.55}$,    
which is the best fit to the data of \citet{SchTil09}. The inset is a close-up of the data of the series~S1--S3 
for $4 \times 10^{-5} \leq \Ra_{f\ast} \leq 10^{-3}$ on the $x$-axis and 
$8\times 10^{-4} \leq \Nu_{\ast} \leq 3\times 10^{-3}$ on the $y$-axis.}
\label{fig:Nus_Rafs}
\end{figure}

Finally, we assess if the reduction of the Nusselt number due to the presence of the LSV affects 
the scaling law deduced for heat flux measurements in the numerical study of \citet{SchTil09}. 
We use the results of \citeauthor{SchTil09} for comparison, since their rotating Rayleigh-B\'enard convection 
model is similar to ours, with fixed temperature and  stress-free boundary conditions, although they consider 
Prandtl numbers of $\Pran=7$ and $\Pran=0.7$. To obtain a scaling law for the heat flux that is independent 
of the diffusivities $\kappa$ and $\nu$, they seek a scaling of the form
\begin{equation}
	\Nu_{\ast} = \alpha \Ra_{f\ast}^{\beta},
	\label{eq:scaling_tilgner}
\end{equation}
with $\Nu_{\ast} = \Nu\Ek/\Pran$ and the flux Rayleigh number $\Ra_{f\ast} = \Ra \Ek^3 \Nu / \Pran^2$. 
\citet{SchTil09} find that the best fit to their data is obtained for $\alpha=0.17$
and $\beta=0.55$. 
The prefactor $\alpha$ given here takes into account the different definitions of $\Ek$ used
in \citeauthor{SchTil09} and in this paper.
It should be noted that they obtained this scaling based on the data points for which $0.5< \Rey \Pran \Ek^{1/2} < 10$. 
As shown previously in figure~\ref{fig:ReRez_tilgner}, most of our data points are indeed within this interval 
(replacing $\Rey$ in \citeauthor{SchTil09} by $\sqrt{3}\Rey_z$ in our simulations). 
In figure~\ref{fig:Nus_Rafs}, we plot $\Nu_{\ast}$ as a function of $\Ra_{f\ast}$ in our simulations. 
As observed in \citeauthor{SchTil09}, no individual series follows the scaling~(\ref{eq:scaling_tilgner}) 
particularly well, but the exponent $\beta=0.55$ is a good fit to the envelope defined by all of the data points. 

The reduction of $\Nu_{\ast}$ for $\aspect=4$ (series~S3) and $\aspect=2$ (S2) compared with $\aspect=1$ (S1) is 
barely visible on the $y$-axis of figure~\ref{fig:Nus_Rafs} since it spans four decades. The inset shows a close-up of 
the data of the series~S1--S3 over less than a decade of the $y$-axis. Whereas the reduction of $\Nu_{\ast}$ is visible 
on the inset, it remains small compared with the variation of $\Nu_{\ast}$ with $\Ra_{f\ast}$. In practice, this means 
that the effect on the scaling laws for the heat flux of changing $\lambda$, \ie changing the amplitude of the LSV,  is 
relatively small.

\section{Discussion}
\label{sec:discussion}

We have presented simulations of rotating Rayleigh-B\'enard (RRB) convection that demonstrate the emergence of 
long-lived, large-scale vortices (LSV). These LSV consist of a patch of strong 
cyclonic vorticity surrounded by a region of weaker anticyclonic vorticity, both aligned with the rotation axis, 
which appear at the box size
and are nearly depth-independent. With stress-free top and bottom boundaries, for the Ekman numbers considered 
here \mbox{($\Ek=10^{-4}$ -- $5\times 10^{-6}$)}  and depending on the aspect ratio, the kinetic energy of the 
horizontal flow can be as much as ten times greater than that of the vertical flow, which is driven directly by buoyancy. 
LSV are observed when the Reynolds number based on the rms vertical velocity exceeds $100 \textrm{ -- } 300$, the 
threshold value being dependent on the box aspect ratio but independent of the Ekman number. This corresponds to Rayleigh 
numbers only about three times that at the onset of convection. 
The amplitude of the large-scale flow starts to decline once the thermal input is strong enough to allow a relaxation 
of the rotational constraint. Quantitatively, this decay of the LSV occurs for a local Rossby number based on the 
convective velocity, $\Ro_z^l$, of approximately $0.15$. 
Moreover, if the two conditions (i) \mbox{$\Rey_z > 100 \textrm{ -- } 300$} and (ii) \mbox{$\Ro_z^l \lesssim 0.15$} 
are met, we always observe a transfer of energy to the large 
horizontal scale, even for modest scale separation between the horizontal convective eddies and the horizontal extent of 
the domain (a factor four between the two is the smallest scale separation we considered). We tested the cyclone/anticyclone 
asymmetry of the LSV by artificially inverting the sign of the vorticity at a given time; the large-scale anticyclone 
subsequently disintegrates into smaller vortices, and the cyclone/anticyclone asymmetry at large scales is established 
relatively rapidly, after about $100$ rotation timescales.

To gain some insight into the mechanism of the formation of the LSV, we performed a series of filtered simulations, 
in which spectral coefficients of given horizontal and vertical wavenumbers, $k_x$, $k_y$ and $k_z$, 
are artificially suppressed during the time integration. The filtered simulations suggest that the LSV 
(corresponding to $(k_x,k_y,k_z)=(1,1,0)$ in spectral space)
are produced by the nonlinear interactions of small-scale $z$-dependent convective motions. Moreover, the presence of 
the spectral range between $(k_x,k_y)=(1,1)$ 
and the typical horizontal wavenumber of the convective structures is not required to sustain the LSV. 
As mentioned above, the amplitude of the LSV declines if the convection is not strongly influenced by rotation, 
in which case the convective structures are less anisotropic. 
To interact coherently, 
the convective motions must therefore present a significant anisotropy between 
their vertical and horizontal extents, \ie they must be significantly affected by rotation.

In our study, the smallest compensated Rayleigh number, $\tRa$, at which LSV appear for different Ekman numbers 
corresponds to the transition from cellular convection ($\tRa \lesssim 20$) to the thermal plumes ($\tRa \gtrsim 20$)
measured in the study of \citet{Jul12}, which is based on a reduced model of Boussinesq convection valid in 
the small Rossby number limit. 
Thermal plumes originate from a buoyancy instability in the thermal boundary layers.
Vortex stretching within the plumes ejected from the boundaries yields
an axial vorticity distribution  
that is skewed towards positive 
values near the top and bottom boundaries \citep{Chen1989,Julien1996}. Since the axial vorticity has no horizontal average, 
anticyclonic convective structures are necessarily also present, but they are less compact and have weaker vorticity. 
The formation of intense cyclonic thermal plumes near the boundaries could explain the predominance of the large-scale 
cyclonic circulation. Assuming that two cyclonic plumes form from the thermal boundary layer at sufficiently small distance, 
they would start to drift horizontally around one another \citep[e.g.][]{Boubnov1986,Hopfinger1993}; the conditions 
for the merger of two like-signed vortices depend notably on their separation distance, radius, and vorticity, and 
are the subject of an abundant literature on vortex dynamics 
\citep[e.g.][]{Griffiths1987, Melander1988, Cerretelli2003, Meunier2005}. 
The patch of cyclonic vorticity they create will be deformed by the background shear created by nearby individual vortices. 
In return, the deformed cyclonic patch tends to attract nearby cyclones and repel anticyclones \citep{Yasuda1997}. 
Merging of anticyclonic structures can also occur, but would be less likely as vortices of intense strength are more 
likely to merge. Since the horizontal boundaries are periodic, the repulsion of anticyclones by the large-scale cyclone 
would tend to group the anticyclones in the surrounding area, and so establish the weak anticyclonic circulation. 
In this scenario, the underlying asymmetry between large-scale cyclonic and anticyclonic circulation arises therefore 
through the formation of thermal plumes, which builds up a population of strong narrow cyclonic vortices, and the 
interactions between like-signed vortices then lead to the formation of one large cyclonic vortex by absorbing this 
available population of strong narrow cyclones. A potential weakness of this explanation for LSV formation is that 
the initial population of narrow cyclones is mainly located near the horizontal boundaries, whereas the LSV span the 
entire vertical extent of the domain. The clustering of cyclonic vorticity is described here as a two-dimensional process, 
but the conditions for interaction and merger of three-dimensional vortices have also been studied in detail in the 
literature \citep[e.g.][]{Ozuugurlu2008}. As observed during the time evolution of the kinetic energy of 
the large 
horizontal scale, the process of formation is slow, and occurs over thousands of rotation timescales; 
the large-scale flow eventually saturates when the viscous dissipation can balance 
the clustering of the convective eddies.

In their reduced model, \citet{Jul12} observe the thermal plume regime for $\tRa \gtrsim 20$, but they report 
the formation of LSV for larger Rayleigh number, namely $\tRa=100$. In their study, the large-scale depth-invariant 
mode consists of a cyclone/anticyclone pair of similar vorticity. Towards the low Rossby number limit, 
the asymmetry between cyclonic and anticyclonic thermal plumes tends to vanish \citep{Vorobieff2002,Sprague2006}, 
and we indeed expect that the process of clustering of like-sign vorticity plumes would produce both large-scale cyclonic 
and anticyclonic circulation of equal strength.

The simple scenario we propose for the formation of the large-scale cyclone is in agreement with the result of our 
filtered simulations, but remains to some degree speculative. This proposed picture could work in conjunction with 
the instability of large-scale anticyclonic regions for which the total vorticity ($\omega_z + 2 \Omega$) is 
small \citep{Lesieur91}.
The reduction of the rotational constraint on the convection in large-scale anticyclonic 
regions yielding a possible local increase of $\Ro_z^l$ above the threshold value of $0.15$
could also contribute to the preference for cyclonic LSV. 
To confirm the proposed 
scenario of the formation of LSV, it would be interesting to study the interactions of a small number of convective 
structures in isolation, together with the effect of artificially added large patches of vorticity. Such studies are 
beyond the scope of the present paper, but could be addressed in future work.

Despite a number of fully 3D models of rotating Boussinesq convection in Cartesian boxes, this work is one of the first 
to report the formation of box-size vortices in this system \citep[see also][]{Favier2014}. The independent work 
of \citet{Favier2014} has been carried out using a model of RRB convection with the same boundary conditions as ours, 
and their results are in agreement concerning the domain of existence of the LSV, the nature of the energy transfer 
to large scales, and the asymmetry between cyclones and anticyclones. An interesting difference between the two 
studies is that \citeauthor{Favier2014} use a computational domain of larger aspect ratio for similar Ekman numbers 
(for instance $\lambda=4$ for $\Ek=10^{-5}$), so they were able to achieve larger scale separation between the box size 
and the convective scales. In this case, they observe that several coherent cyclonic vortices coexist initially, and 
that these merge when two of these cyclones become close together, with eventually only one box-size cyclone remaining.

Some of the earlier numerical studies of RRB convection have been carried out in the same parameter regime for which 
we identified the existence of LSV. Most of these were interested in measuring the heat flux in order to deduce scaling 
laws and thus identify transitions between the various convection regimes. As shown in \S\,\ref{sec:heat}, the presence 
of LSV markedly disturbs the convection by inhibiting the mixing in the core of the cyclone, yielding a reduction of 
the Nusselt number compared with its value when convection sets in, of about $5$ to $8$\%, depending on the aspect ratio. 
However, when viewed over several decades of the input parameters, this effect on the Nusselt number is not particularly 
noticeable on the scaling laws calculated with different aspect ratios. LSV could therefore be present in these earlier 
studies but not reported because of their minor influence on the scaling laws of the heat flux. 

The choice of boundary conditions is probably an important factor for the formation of LSV.  No-slip boundary conditions 
for the velocity would tend to increase the viscous damping in the boundary layers compared with stress-free conditions, 
thereby reducing the amplitude of the horizontal flows. 
The absence of LSV in a number of experimental studies conducted in the range of parameters where LSV might 
be expected \citep[e.g.][]{Boubnov1986,Zhong09,King2012b} possibly suggests a destructive effect of no-slip 
boundaries on these structures. However, the comparison between simulations and experiments is not entirely straightforward 
because the main difference lies not only in the top and bottom boundary conditions, but also in the presence of side walls, 
which are known to influence convection in some cases \citep{Liu99}. Furthermore, in order to observe LSV in fluids with low 
viscosity, it is necessary to run experiments for a long time since the saturation depends on the viscous dissipation at 
large scales. Finally, changing the boundary conditions for the temperature to fixed flux rather than fixed temperature 
may also affect the presence of LSV, although the effect in this case is more difficult to predict. 
The effect of the boundary conditions therefore remains an interesting open question, which we propose 
to investigate in subsequent work.

\section*{Acknowledgements}
This work was supported by the Natural Environment Research Council under grant NE/J007080/1. 
This work was undertaken on ARC1 and ARC2, part of the High Performance Computing facilities at the University
of Leeds. This work also used the COSMA Data Centric system at Durham University, operated by the Institute for
Computational Cosmology on behalf of the STFC DiRAC HPC Facility (www.dirac.ac.uk). This equipment was 
funded by a BIS National E-infrastructure capital grant ST/K00042X/1, DiRAC Operations grant ST/K003267/1 and 
Durham University. DiRAC is part of the National E-Infrastructure.
We are grateful to Toby Wood and Benjamin Favier for 
helpful discussions and to two anonymous referees for suggestions that have improved the manuscript.

\section*{Supplementary movies}
Supplementary movies are available at \url{http://dx.doi.org/10.1017/jfm.2014.542}.

\end{document}